\documentclass[nofootinbib,twocolumn,pra,superscriptaddress]{revtex4-2}
\usepackage{graphicx}%
\usepackage{dcolumn}%
\usepackage{bm}%
\usepackage{float}
\usepackage[usenames,dvipsnames]{color} %
\usepackage{xcolor}
\usepackage{hyperref}
\hypersetup{colorlinks=true, linkcolor=blue!80!black, urlcolor=blue!80!black, citecolor=blue!80!black}

\usepackage{CJK}
\usepackage{comment}
\usepackage{adjustbox}
\usepackage{appendix}
\usepackage{braket}
\usepackage{xfrac}
\usepackage{multirow, makecell}
\usepackage{enumitem}
\usepackage{yquant}
\useyquantlanguage{qasm}
\usetikzlibrary{quantikz2}

\newcommand{\Gate}[1]{\textsc{#1}}

\newcommand{\cnotgate}{\Gate{cnot}}
\newcommand{\cxgate}{\Gate{cnot}}

\newcommand{\rev}[1]{\textcolor{black}{#1}}

\newtheorem{simplification}{Simplification}

\begin{document}

\preprint{APS/123-QED}

\title{Performance of Quantum Approximate Optimization \\ with Quantum Error Detection}%

\author{Zichang He}
\thanks{These authors contributed equally to this work. Correspondence should be addressed to \texttt{zichang.he@jpmchase.com}.}
\affiliation{%
 Global Technology Applied Research, JPMorganChase, New York, NY 10001, USA
}%

\author{David Amaro}
\thanks{These authors contributed equally to this work. Correspondence should be addressed to \texttt{zichang.he@jpmchase.com}.}
\affiliation{%
Quantinuum, Partnership House, Carlisle Place, London SW1P 1BX, United Kingdom
}%
\author{Ruslan Shaydulin}
\affiliation{%
 Global Technology Applied Research, JPMorganChase, New York, NY 10001, USA
}%
\author{Marco Pistoia}
\affiliation{%
 Global Technology Applied Research, JPMorganChase, New York, NY 10001, USA
}%

\date{\today}

\begin{abstract}
Quantum algorithms must be scaled up to tackle real-world applications. Doing so requires overcoming the noise present on today's hardware. The quantum approximate optimization algorithm (QAOA) is a promising candidate for scaling up, due to its modest resource requirements and documented asymptotic speedup over state-of-the-art classical algorithms for some problems. However, achieving better-than-classical performance with QAOA is believed to require fault tolerance. In this paper, we demonstrate a partially fault-tolerant implementation of QAOA using the $[[k+2,k,2]]$ ``Iceberg'' error detection code. We observe that encoding the circuit with the Iceberg code improves the algorithmic performance as compared to the unencoded circuit for problems with up to $20$ logical qubits on a trapped-ion quantum computer. Additionally, we propose and calibrate a model for predicting the code performance. We use this model to characterize the limits of the Iceberg code and extrapolate its performance to future hardware with improved error rates. In particular, we show how our model can be used to determine the necessary conditions for QAOA to outperform the Goemans-Williamson algorithm on future hardware. To the best of our knowledge, our results demonstrate the largest universal quantum computing algorithm protected by partially fault-tolerant quantum error detection on practical applications to date, paving the way towards solving real-world applications with quantum computers. 
\end{abstract}

\maketitle

\section{Introduction}
Quantum computers are poised to deliver algorithmic speedups for a broad range of applications in science and industry~\cite{2310.03011,herman2023quantum,DAC24_review,chakrabarti2024generalized}. However, realizing these speedups requires overcoming the challenge presented by the noise which limits the computational power of today's quantum devices. Error correction~\cite{gottesman2016surviving} provides a scalable path to fault-tolerance and has shown significant progress in hardware recently~\cite{acharya2023,acharya2024quantum,Xu2024,PhysRevX.11.041058,Bluvstein2023,Gupta2024,Campbell2024,PRXQuantum.5.030326,sivak2023real,reichardt2024}. Nonetheless, quantum error-correction imposes large overheads, making it challenging to execute even small-scale applications fully fault-tolerantly. As a result, fully fault-tolerant demonstrations of quantum algorithms for practical applications have been out of reach of experiments, despite the immense progress in implementation and benchmarking of algorithmic components, such as preparation of magic states~\cite{2310.12106}, one-bit addition~\cite{wang2023fault}, and quantum Fourier transform~\cite{mayer2024benchmarking}.

Quantum error-detection (QED) %
codes provide an opportunity for partially fault-tolerant implementation of algorithms in the near-term~\cite{Bluvstein2023,Gupta2024,self2024protecting,hangleiter2024fault,2404.02280,gerhard2024weakly,Linke2017}. While non-scalable, they can still deliver improved algorithmic performance beyond what is possible without protecting against noise~\cite{Pokharel2024,gowrishankar2024logical,yamamoto2024demonstrating,nishi2024encoded}. The protection offered by QED codes opens an opportunity to use quantum computers to study the performance of quantum algorithms for sizes and noise rates beyond classical simulation.

The recently proposed $[[k+2,k,2]]$ ``Iceberg'' QED code~\cite{self2024protecting} is particularly suitable for near-term algorithms due to its ability to encode expressive circuits, using a universal set of local and global logical rotations, with a low overhead. The Iceberg code has been demonstrated to improve the fidelity of random circuits with up to $8$ logical qubits and $1,323$ physical two-qubit gates~\cite{self2024protecting}, the performance of quantum phase estimation with up to $4$ logical qubits and $920$ two-qubit gates~\cite{yamamoto2024demonstrating}, and the fidelity of ground state preparation with probabilistic imaginary-time evolution with $4$ logical qubits and up to $906$ two-qubit gates~\cite{nishi2024encoded}. 
While not fully fault-tolerant, these experiments provided preliminary evidence that for circuits with small numbers of qubits, the Iceberg code can improve algorithmic performance.

Quantum Approximate Optimization Algorithm (QAOA)~\cite{Hogg2000,farhi2014quantum} is a quantum optimization heuristic applicable to a broad range of combinatorial optimization problems in finance and other industries~\cite{hadfield2019quantum,kremenetski2021quantum,soloviev2022quantum,blekos2024review}. QAOA has been shown to provide a quantum algorithmic speedup over state-of-the-art solvers for some problems~\cite{shaydulin2023evidence,2208.06909}, motivating its implementation on hardware. 
While relatively low resource requirements enabled QAOA execution on non-error-corrected hardware~\cite{qaoa_np,pelofske2024short,decross2023qubit,moses2023race,Harrigan2021,heAlignmentInitialState2023,Niroula2022,hao2024end,he2025non}, realizing the speedup offered by QAOA is widely believed to require fault tolerance~\cite{StilckFrana2021,DePalma2023,omanakuttan2025threshold}. 
We remark that both quantum hardware performance and the impact of quantum noise on QAOA have been subjects of extensive interest~\cite{baldwin2022re,sharma2020noise,dalzell2024random,marshall2020characterizing, gonzalez2022error, xue2021effects, he2024_dro, Kakkar2022}.

\begin{figure*}[t]
    \centering
    \includegraphics[width=\linewidth]{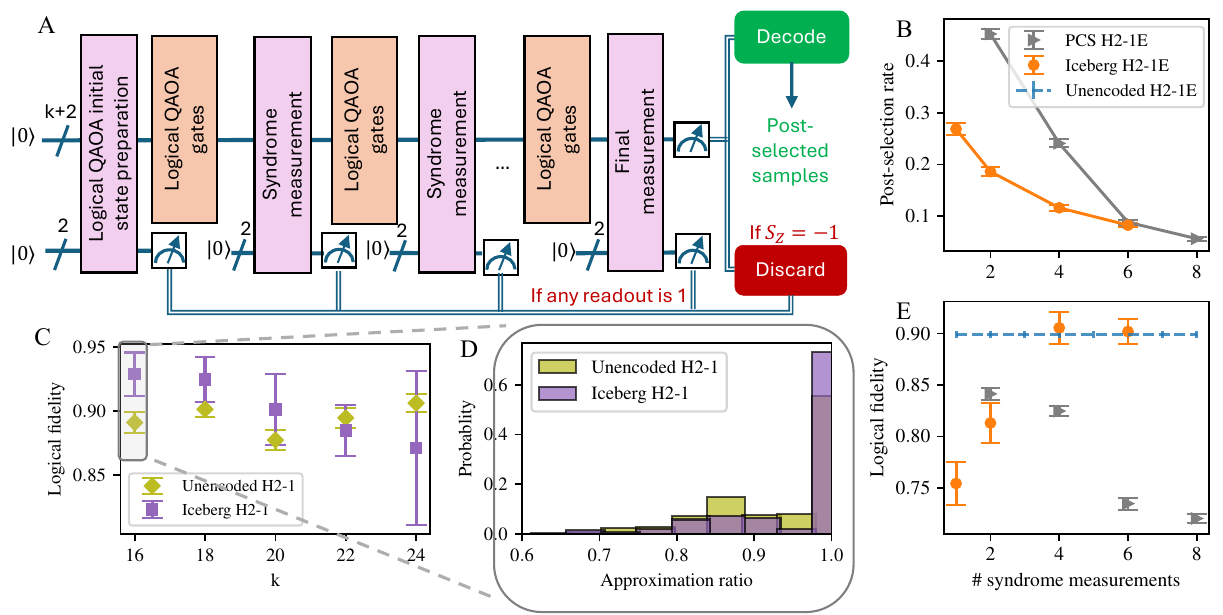}
    \caption{\textbf{Motivation: The Iceberg code is a performant method for error detection in the near-term.} 
    \textbf{A} The Iceberg code detects errors that occur in the execution of a $k$-qubit circuit by encoding it in $(k+2)$ physical qubits. \textbf{B} Shots containing a detected error are discarded, resulting in a post-selection overhead.
    \textbf{C} Performance of QAOA \rev{with 10 layers} with and without the Iceberg code on the Quantinuum H2-1 quantum computer. Here, the logical fidelity (defined in Eq.~\eqref{eq:noisy_ar}) directly indicates the approximation ratio. The Iceberg code \rev{with 4 syndrome measurements} improves performance on small problems, while being detrimental on larger ones. 
    \textbf{D} An example of measured samples with and without the Iceberg code \rev{obtained from the H2-1 device}. After detecting the errors, the probability of the higher energy states is amplified, reflecting the approximation to the noiseless QAOA performance.  
    \textbf{E} The Iceberg code performs better than other commonly-used techniques for \rev{error detection} in QAOA circuits like Pauli Check Sandwiching (PCS)~\cite{gonzales2023quantum}. \rev{Data are obtained from the H2-1 emulator.}
    Error bars show the standard errors. \rev{All hardware data are labeled as H2-1, while all emulated data are labeled as H2-1E.}
    }
    \label{fig:panel-1}
\end{figure*}
We demonstrate a partially fault-tolerant implementation of QAOA applied to the MaxCut problem on the Quantinuum H2-1 trapped-ion quantum computer~\cite{H2} with the Iceberg code. %
At the time of our experiments, the H2-1 device had $32$ all-to-all connected qubits and 99.8\% two-qubit gate fidelity~\cite{moses2023race}.
We execute circuits with up to $24$ logical qubits encoded into up to $26$ physical qubits using up to $813$ physical two-qubit gates.
With this, we observe that protecting the circuit with the Iceberg code leads to an improved approximation ratio, as compared to the unencoded circuit for problems requiring up to $20$ logical qubits. However, we also observe that beyond $20$ logical qubits, 
the Iceberg code does not yield an improved algorithmic performance.
To the best of our knowledge, these experiments are the largest evaluation of the Iceberg code and the largest QED-encoded application demonstration in terms of the number of logical qubits to date. 

To understand the protection capability of the Iceberg code, we propose a model that predicts the code performance. 
The model efficiently constructs an analytical estimation of the logical fidelity and post-selection rate as a function of the circuit size and three error rates related to the noise produced by two-qubit physical gates.
We calibrate the model by simulating a large set of QAOA circuits of varying sizes, with and without the Iceberg code in the emulator~\cite{crathesis, pecos, PhysRevX.11.041058} of the H2-1 quantum computer. The calibrated model is then used to characterize the regimes in which the Iceberg code improves the algorithmic performance of QAOA. 
Specifically, we identify the ranges for the number of logical qubits, QAOA depth, and the number of syndrome measurements for which the Iceberg code is beneficial. Furthermore, our model can be used to predict the performance of the Iceberg code on future improved hardware.
We demonstrate conditions on effective error rates for 
QAOA to outperform the Goemans-Williamson (GW)~\cite{goemans1995improved} algorithm on small graphs. 

\section{Background}
We begin with a brief review of relevant concepts about QAOA and the Iceberg code.

While our model can be generalized to any optimization problem, in this paper we focus on the MaxCut problem as a commonly-studied benchmark problem.

Given a graph $G(k,E)$ with $k$ vertices and set of edges $E$, the MaxCut problem consists in finding a cut that partitions the vertices into two sets that maximize the number of edges between them. Cuts can be represented by strings $\bm{z}$ of $k$ bits with value $z_i=\pm1$ if vertex $i$ is in one set or the other. The MaxCut objective function can be written as 
$f(z) = \sum_{(i,j)\in E} (1 - z_i z_j$).

On qubits, the MaxCut problem is equivalent to finding the ground state of the following $k$-qubit Hamiltonian:
\begin{equation}\label{eq:max_cut}
    \mathcal{H} = \sum_{(i,j) \in E} Z_i Z_j,
\end{equation}
where we define the Pauli operators as $\mathcal{P} = \{I, X, Y, Z\}$ and $Z_i$ as the Pauli-$Z$ operator acting on qubit $i$. The $k$-qubit computational state $\ket{\bm{z}}=\otimes_{i=1}^k \ket{z_i}$, with $Z_i\ket{\bm{z}} = z_i \ket{\bm{z}}$, that minimizes the cost Hamiltonian represents the optimal solution of the problem.%

\subsection{Quantum Approximate Optimization Algorithm}
QAOA is a quantum algorithm for combinatorial optimization.
It solves optimization problems by preparing a quantum state using a sequence of $\ell$ layers of alternating cost Hamiltonian and mixing Hamiltonian operators, parameterized by vectors $\bm\gamma$ and $\bm\beta$, respectively. 
\begin{equation}
    \ket{\bm{\psi}} = e^{-i\beta_\ell\mathcal{M}}e^{-i\gamma_\ell\mathcal{H}} \cdots e^{-i\beta_1\mathcal{M}}e^{-i\gamma_1\mathcal{H}} \ket{\bm{\psi}_0} 
\end{equation}
The parameters $\bm\gamma$ and $\bm\beta$ are chosen such that the measurement outcomes of $\ket{\bm{\psi}}$ correspond to high-quality solutions of the optimization problem with high probability. In this paper, we take the initial state $\ket{\bm{\psi}_0} = \ket{+}^{\otimes k}$ as the equal superposition of all possible candidate solutions, and the mixing Hamiltonian as a summation of all single-qubit Pauli-$X$ operators $\mathcal{M} = \sum_{i=1}^n X_i$. 

Denoting the value of optimal cut by $f_{\rm max}$, we can quantify how well QAOA with state $\ket{\bm{\psi}}$ solves the MaxCut problem by computing the \textit{approximation ratio}:
\begin{equation}
  \alpha(\bm{\psi}) = \frac{|E| - \braket{\bm{\psi}|\mathcal{H}|\bm{\psi}}}{2 f_{\rm max}}.
\end{equation}

Recent progress in parameter setting heuristics has considerably advanced the execution of QAOA in the early fault-tolerant era~\cite{ICCAD_qaoapara,hao2024end}, with good parameter choices available for many problems.
A set of parameters that leads to good approximation ratios was proposed in~\cite{Wurtz2021} for the MaxCut problems on regular graphs that we solve in this work. Throughout our paper, we use these ``fixed angles'' to set QAOA parameters in the experiments.

\subsection{Iceberg Code}
The Iceberg code protects $k$ (even) logical qubits with $n=k+2$ physical qubits and two ancillary qubits. We label the physical qubits as $\{t, 1, 2, \ldots, k, b\}$, where the two additional qubits are called \textit{top} $t$ and \textit{bottom} $b$ for convenience. The two code stabilizers and the logical operators are
\begin{eqnarray}
    S_X &&= X_t X_b \prod_{i=1}^k X_i, \\
    S_Z &&= Z_t Z_b \prod_{i=1}^k Z_i, \\
    \bar{X}_i &&= X_t X_i \quad \forall \, i\in\{1, 2, \ldots, k\}, \\
    \bar{Z}_i &&= Z_b Z_i \quad \forall \, i\in\{1, 2, \ldots, k\}.
\end{eqnarray}
From these definitions one can see that the logical gates of the QAOA circuit are implemented as the physical gates
\begin{eqnarray}
    \exp(-i\beta\bar{X}_i) = \exp(-i\beta X_tX_i), \label{eq:sq_Xgate}\\
    \exp(-i\gamma\bar{Z}_i\bar{Z}_j) = \exp(-i\gamma Z_iZ_j). \label{eq:tq_Zgate}
\end{eqnarray}
In Quantinuum devices, these physical gates are implemented by just one native two-qubit gate $\exp(-i\theta Z_iZ_j)$ and various single-qubit Clifford gates. 

As depicted in Fig.~\ref{fig:panel-1}A, the Iceberg code employs an initialization block to prepare the initial QAOA state $\ket{\bar{+}}^{\otimes n}$ in the common $+1$ eigenspace of the stabilizers. The logical QAOA gates in Eqs.~\eqref{eq:sq_Xgate} and \eqref{eq:tq_Zgate} are then implemented in blocks, interleaved with syndrome measurement blocks until the QAOA circuit is complete. These syndrome measurement blocks measure the stabilizers regularly across the circuit to prevent the accumulation of noise. The final measurement block measures the stabilizers as well as the $k$ data qubits. The precise form of these blocks is depicted in Appendix~\ref{app:iceberg_circuit}. Accepted samples can be decoded by classical post-processing and serve as a candidate solutions for the problem. 

To detect these errors, the fault-tolerant initialization, syndrome measurement, and final measurement blocks employ two ancillas. In the absence of noise, the state remains purely in the $+1$ eigenspace of the stabilizers during the entire circuit execution and ancillas always output a $+1$ when measured. The final measurement block additionally measures the stabilizer $S_Z$,
which is also expected to be measured as $+1$ in the absence of noise. Therefore, a $-1$ output in any of them signals the presence of an error caused by noise, and the circuit execution is 
discarded. 

The fault-tolerant design of the initialization, syndrome measurement, and final measurement blocks ensures that no single faulty component in these blocks (like a two-qubit gate) can cause a logical error. In contrast, our logical gates, despite being natural for the hardware, are not fault-tolerant, as some errors in their physical implementation cannot be detected. Nevertheless, we show in Sec.~\ref{sssec:fitted_err_rates} that undetectable errors are rare, rendering the QAOA protection of the Iceberg code effectively fully fault-tolerant. 

\section{Results}

We now present our results. First, we summarize the results obtained on the hardware. Then, we discuss the model fitting results and the performance predictions on future hardware.
The H-series emulators~\cite{crathesis, pecos, PhysRevX.11.041058} we use perform a state-vector simulation where noise is randomly sampled following realistic noise models and then inserted into the circuit. Currently, the most influential noise channels are gate errors and single-qubit coherent dephasing from memory errors. The performance gap between the hardware and emulator experiments is discussed in Appendix~\ref{app:data} for completeness. 

\begin{figure*}[t]
    \centering
    \includegraphics[width=\linewidth]{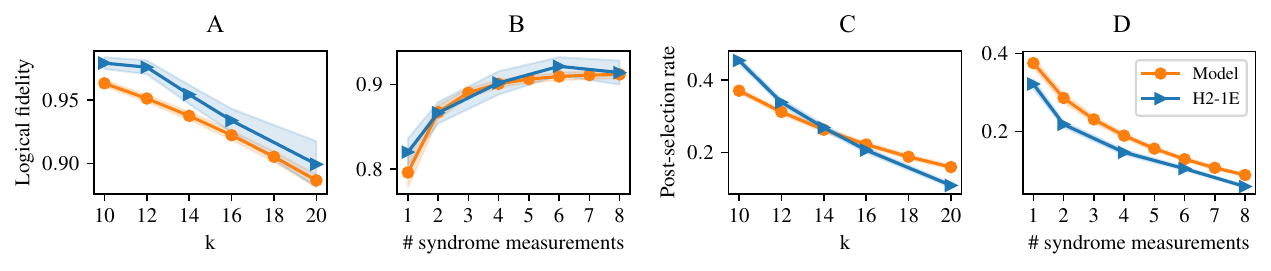}
    \caption{\textbf{Proposed model accurately reflects behavior of Iceberg encoding circuits observed in high-accuracy emulation.} The fitted model matches the qualitative and quantitative behavior of logical fidelity and post-selection rate for both varying qubit count with a fixed $\ell=9$ (\textbf{A,C}) and varying number of syndrome measurements with a fixed $k=16$ and $\ell=11$ (\textbf{B,D}). The shaded regions represent the standard errors.
    }
    \label{fig:panel-2}
\end{figure*}

\subsection{Iceberg code protection of QAOA on hardware}
The performance of the Iceberg code with QAOA on 3-regular graph MaxCut on the Quantinuum H2-1 quantum computer~\cite{H2} is shown in Fig.~\ref{fig:panel-1}C.  
The logical fidelity reported in this figure is estimated from the average energy measured experimentally by assuming a global white noise model distribution, as described in Sec.~\ref{ssec:white_noise_model}.
We fix the QAOA depth to $\ell=10$ and vary the number of logical qubits $k$, randomly selecting one MaxCut graph instance per $k$. For each problem we run QAOA unencoded and QAOA protected by the Iceberg code with three intermediate syndrome measurements. 
Throughout this paper, the final measurement is counted as a syndrome measurement, so in the previous experiment we say that four syndrome measurements are used. The Iceberg data has larger error bars due to the smaller number of post-selected samples.
The performance of Iceberg circuits decreases as $k$ increases. Meanwhile, the performance of unencoded circuits is relatively robust because the accumulated error of additional gates when increasing $k$ is not significant to diminish performance at this scale.

The histogram in Fig.~\ref{fig:panel-1}D reports the hardware shots of several Iceberg and unencoded QAOA circuits for $k=16$. After post-selection, the output distribution has higher weight on bitstrings with a higher approximation ratio, as expected from a better protection against noise.

We compare the performance with that of the Pauli check sandwiching (PCS)~\cite{gonzales2023quantum,PhysRevResearch_CPC}, an error detection scheme with a similar motivation to that of the Iceberg code. PCS uses pairs of parity checks to detect some but not necessarily all errors that occur in a given part of the circuit. The parity checks are chosen based on the symmetries already present in the circuit. For QAOA circuits considered in this work, the problem Hamiltonian commutes with $X^{\otimes k}$ and $Z^{\otimes k}$, so we use them as the checks of our PCS experiments. 
To unify the notation with Iceberg code, a pair of $X^{\otimes k}$ and $Z^{\otimes k}$ checks is denoted as one syndrome measurement. For example, one syndrome measurement in PCS means that we select one cost Hamiltonian layer $e^{-i \gamma \mathcal{H}}$ and sandwich it with two parity checks. The overhead of one syndrome measurement in PCS includes an additional $4k$ two-qubit gates, along with two ancillas. The comparison with PCS on a $k=18$, $\ell=11$ QAOA circuit is shown in Fig.~\ref{fig:panel-1}B and E, where all data are from the H2-1 emulator (H2-1E). 
We observe that PCS leads to a lower logical fidelity that does not increase with the number of syndrome measurements.
At this scale, the large overhead and the non-fault tolerant design of the PCS method decreases the circuit performance.
At the same time, we observe that the Iceberg code can effectively improve the QAOA performance and obtain a higher logical fidelity than the unencoded circuit with four syndrome measurements. 
\rev{Here, we do not aim to claim QED is better than quantum error mitigation techniques as QEM usually aims to improve the accuracy of estimating an observable, rather than improving the quality of a single bitstring outcome. 
This study does not aim to claim the superiority of QED over quantum error mitigation (QEM) techniques, as QEM is generally designed to enhance the accuracy of observable estimations rather than the quality of individual bitstring outcomes. PCS was chosen as the baseline due to its unique capability among QEM methods to perform error detection, which aligns with the principles of Iceberg encoding, albeit without the necessity of encoding.}

\subsection{Estimated performance from our model}
To understand the protection capability of Iceberg code, we propose the performance model of Sec.~\ref{ssec:model} for the unencoded and Iceberg code circuits. The model outputs analytical functions of the logical fidelity $\mathcal{F}_\mathrm{une}$ for the unencoded circuits, the logical fidelity $\mathcal{F}_\mathrm{ice}$ for the Iceberg code circuits, and the post-selection rate $1-D$ for the Iceberg code ($D$ is the discard rate). Inputs from the circuit are the numbers of logical qubits $k$, logical single-qubit gates $g_1$, logical two-qubit gates $g_2$, and syndrome measurements $s$. From the hardware, the model for the unencoded and the Iceberg code respectively inputs only one and three error rates related to the noise produced by two-qubit physical gates. These are motivated and described in more detail in the next section. We leave the error rates as fitting parameters so that, when fitted to data from the H2-1 emulator, the fitted values incorporate corrections from other noise sources.

Since QAOA experiments on hardware or emulators output the approximation ratio instead of the logical fidelity, we extend the performance model by approximating the noise distribution as that of a global white noise. 
Details are discussed in Sec.~\ref{ssec:white_noise_model}.   

\begin{table}[b]
    \centering
    \begin{adjustbox}{width=\linewidth}
    \begin{tabular}{c|ccc|c}
        \hline
          & &Iceberg & & Unencoded \\ 
        Parameter & $p_{cx}$ & $p_c$ & $p_a$ & $p_{\ell}$\\ \hline 
        Emulator & 1.28e-3 & [1.3e-05, 3.2e-05] & [4.3e-4, 1.1e-3] & [4.7e-4, 1.0e-3] \\ \hline
        \textbf{Model} & 5.5e-3 & 7.0e-5 & 2.2e-3 & 4.4e-4\\
        \textbf{CI}   & [5.0e-3, 6.2e-3] & [4.3e-5, 9.5e-4] & [1.9e-3, 2.5e-3] & [4.0e-4, 5.0e-4] \\ \hline
        
    \end{tabular}
    \end{adjustbox}
    \caption{Error rates from the H2-1 emulator~\cite{crathesis, pecos, PhysRevX.11.041058} and the performance model.}
    \label{tab:fitted_para}
\end{table}

\subsubsection{Dataset and model validation}
We use the emulator of the Quantinuum H2-1 quantum computer to generate a dataset with varying number of logical qubits in the range $k \in [8, 26]$, QAOA layers in the range $\ell \in [1, 11]$, and syndrome measurements in the range $s\in[1, 8]$. %
This dataset contains 115 circuits for the Iceberg code and 56 for the unencoded circuits. We take $1000$ shots for the unencoded circuits and $3000$ shots for the Iceberg code circuits before post-selection.

From this dataset we select partial data that have relatively stable logical fidelity and large numbers of two-qubit gates to fit the model. For the Iceberg code model, which has three fitting parameters, we use data from $64$ encoded circuits. The data from $15$ unencoded circuits is used to fit the unencoded model, which has only one fitting parameter. We additionally filter out those Hamiltonian terms of every QAOA circuit whose expected values are outliers with respect to the white noise approximation. More details are provided in Appendix~\ref{app:data}. The mean and $95\%$ confidence interval of the fitted parameters from $3000$ boostrapping iterations are reported in Table~\ref{tab:fitted_para}.
\begin{figure*}  
    \centering
    \includegraphics[width=\linewidth]{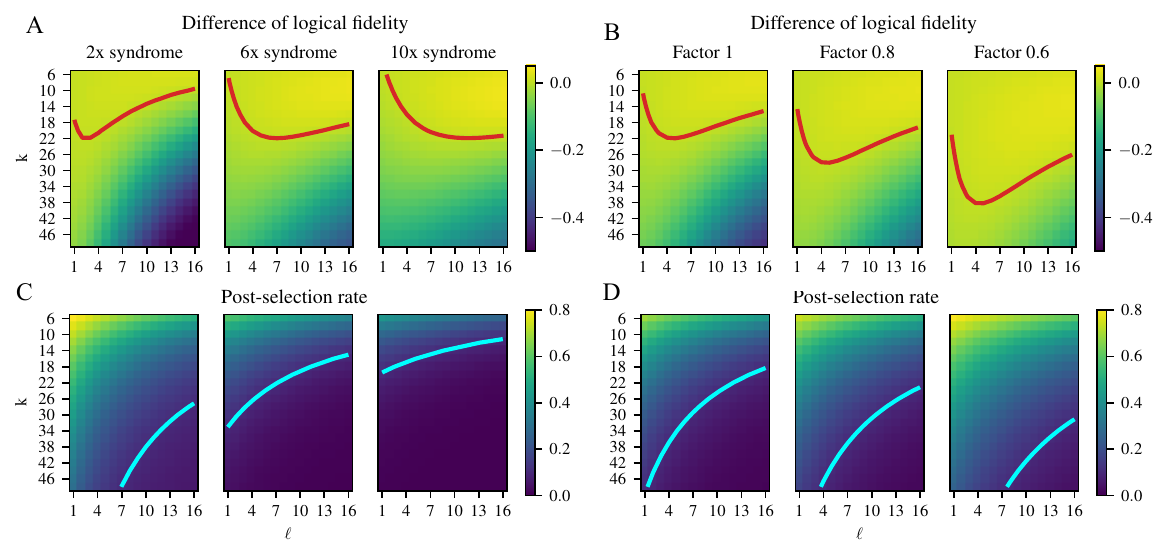}
    \caption{ \textbf{Model prediction}: Predicting the performance of QAOA with the number of logical qubits in the range $k \in [6, 48]$ and the number of QAOA layers in the range $\ell \in [1, 16]$. We use the model proposed in this work to estimate (\textbf{A, B}) the difference $\mathcal{F}_\mathrm{ice} - \mathcal{F}_\mathrm{une}$ in logical fidelity between the Iceberg and the unencoded circuits, and to estimate (\textbf{C, D}) the post-selection rate. In \textbf{A} and \textbf{C}, we use the model fitted error rates in Table~\ref{tab:fitted_para} and vary the number of syndrome measurements. 
    In \textbf{B} and \textbf{D}, we fix the number of syndrome measurements at 4 and scale down the model error rates by the indicated factors.
    The red lines in the top row (\textbf{A}, \textbf{B}) show where the logical fidelity of Iceberg code circuits equals that of unencoded circuits. The cyan lines in the bottom row (\textbf{C}, \textbf{D}) indicate where the Iceberg code circuits have a 10\% post-selection rate.
    }
    \label{fig:panel-3}
\end{figure*}

To validate the accuracy of the performance model, we present the logical fidelity and post-selection rate of \rev{emulated} data alongside the model predictions in Fig.~\ref{fig:panel-2}. We find that our model can match the \rev{emulated} results both qualitatively and quantitatively. The model and experimental fidelities and post-selection rates for every selected circuit, as well as the deviations from the white noise simplification, are presented in Appendix~\ref{app:data}.

\subsubsection{Fitted error rates} \label{sssec:fitted_err_rates}
The deviations between the fitted error rates and the emulator noise rates are presented in Table~\ref{tab:fitted_para}. The error probabilities from the fitted model are larger than the ones from the emulator and provide valuable insights into the accumulation of other noise sources. 

Starting from the error rate $p_{cx}$ of $\cxgate$ gates, the fitted value is more than four times larger than the emulator error rate, showing that a significant amount of noise unaccounted by the performance model accumulates in the error detection blocks of the Iceberg code. 
For the logical gates, we consider two noise channels with error rates $p_c$ and $p_a$ that introduce Pauli errors which commute and anti-commute, respectively, with the two Iceberg code stabilizers. Since the values established in the emulator depend on the rotation angles of the logical gates and the QAOA circuits do not present a clear tendency towards any particular angle, Table~\ref{tab:fitted_para} presents the minimum and maximum values. We find that these fitted error rates are almost double the maximum value given by the emulator, hinting again that the logical gates in the Iceberg code circuit accumulate unaccounted noise. In contrast, the fitted value of the error rate $p_\ell$ of the unencoded logical gates is well approximated by the minimum value established by the emulator.

Importantly, the only single errors that can cause a logical error in the Iceberg code logical gates happen with the smallest probability $p_c \sim $7e-5 among the three noise sources. This indicates that for circuits with a small number of logical gates, the Iceberg code effectively behaves as a fully fault-tolerant quantum error detection code.

The values reported for the emulator in Table~\ref{tab:fitted_para} are obtained from the parameters of the depolarising channel for the native two-qubit gates $\exp(-i\theta ZZ)$ and their dependence with the QAOA rotation angles $\theta \in \{\gamma, \beta \}$. The depolarising channel assigns a probability $q_\sigma$ to each of the 15 Pauli errors in $\sigma = \mathcal{P}^{\otimes2} \setminus \{I^{\otimes2}\}$ after the native gate. The total probability of a Pauli error is the value of $p_{cx}$ reported in the table. These values are corrected by a multiplicative factor defined as a linearly increasing function $r(\theta) \simeq a+b|\theta|$ in the angle magnitude, such that $r(\pi/4)=1$ for maximally entangling gates like the $\cnotgate$. The error rate of unencoded logical gates is then $p_\ell(\theta) = p_{cx}r(\theta)$. 
Additionally, for the Iceberg code, we separate the error rate of commuting errors $q_c = q_{X^{\otimes2}} + q_{Y^{\otimes2}} + q_{Z^{\otimes2}}$ from that of anti-commuting errors $q_a = p_{cx}-q_c$. To unify with the model error rates, we factorize this single channel into a product of a commuting channel with error rate $p_c(\theta) = q_cr(\theta)(1-q_ar(\theta))$ and an anti-commuting channel with error rate $p_a(\theta) = q_ar(\theta)$. We report the minimum and maximum among all QAOA rotation angles.

\subsubsection{Frontiers of the Iceberg Code performance.}
Next, we analyze the performance of the Iceberg QAOA circuits based on the fitted model. We report the difference $\mathcal{F_\mathrm{ice}} - \mathcal{F_\mathrm{une}}$ in logical fidelity between Iceberg and unencoded circuits in Fig.~\ref{fig:panel-3}A, and the post-selection rate of Iceberg circuits $1-D$ in Fig.~\ref{fig:panel-3}C, for varying numbers of syndrome measurements.

As observed from the shift of the breakeven frontiers (red lines) in Fig.~\ref{fig:panel-3}A, the QAOA logical fidelity stabilizes as the number of syndrome measurements increases, even though the post-selection rate, indicated by the cyan lines in Fig.~\ref{fig:panel-3}C, decreases. This aligns with our \rev{emulated} data and findings in the literature~\cite{self2024protecting}, which suggest that the initial syndrome measurements significantly enhance circuit performance, while the marginal gains diminish with an increasing number of syndrome measurements.

\begin{figure*}
    \centering
    \includegraphics[width = 0.8\linewidth]{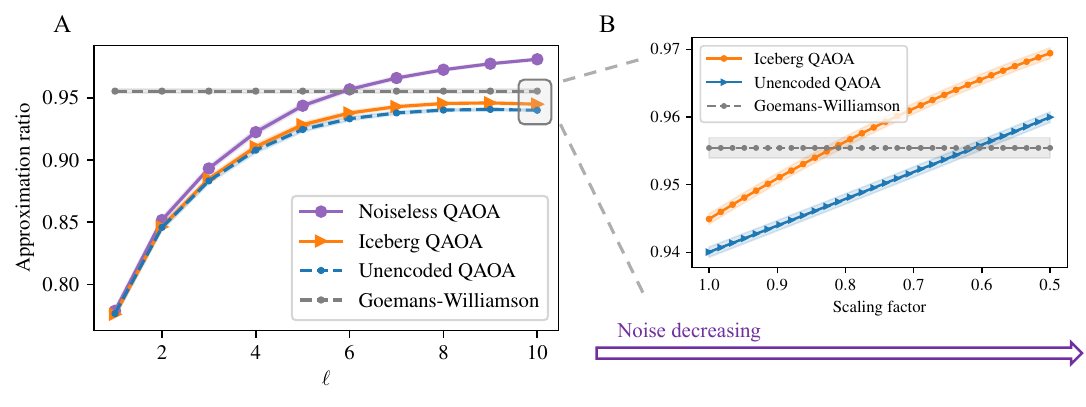}
    \caption{\textbf{Example of using the model to bound the hardware improvement that is necessary but not sufficient for QAOA to become competitive with the Goemans-Williamson algorithm.} 
    \textbf{A} Solve $k=16$ MaxCut using different solvers. Each data is reported as the mean of approximation ratio over 100 $k=16$ $3$-regular graphs. The standard errors are too small to be seen.
    \textbf{B} Scaling of model parameter to beat Goemans-Williamson (GW) algorithm for $k=16$ graphs. The Iceberg code helps the QAOA to beat GW earlier than an unencoded one. \rev{The Iceberg and unencoded QAOA data are obtained from the H2-1 emulator.}
    } 
    \label{fig:scaling_beat_GW}
\end{figure*}
\subsubsection{Predicting performance on future hardware}
We now use our model to predict the performance of QAOA with the Iceberg code on future quantum hardware. To study this, we extrapolate the model performance by scaling all the model parameters in Table~\ref{tab:fitted_para} by a varying factor. A smaller factor corresponds to smaller effective error rates, indicating higher fidelity of the quantum hardware. Scaling all error rates down by the same factor is clearly an additional simplification, as hardware development will not necessarily reduce all noise sources homogeneously and at the same pace. Nevertheless, this analysis provides a valuable qualitative perspective on the potential performance in a foreseeable scenario.

As shown in Fig.~\ref{fig:panel-3}B and D, as the factor decreases, we observe a significant shift of the performance frontier to a larger number $k$ of logical qubits while the post-selection rate improves dramatically. This indicates that with higher-quality quantum hardware, we can push the breakeven frontier of logical fidelity to deeper circuits on larger problem instances with less post-selection overhead.

We can use the model to elucidate the necessary but not sufficient conditions for QAOA to become competitive with classical solvers. As an example, we use our model to answer the question of when a QAOA hardware experiment can outperform the Goemans-Williamson (GW) algorithm~\cite{goemans1995improved} in terms of approximation ratio on small graphs. As reported in the literature~\cite{Wurtz2021}, a noiseless QAOA with fixed parameters has been able to surpass the GW algorithm on small graphs. In Fig.~\ref{fig:scaling_beat_GW}A, we show an example of solving $k=16$-node 3-regular graphs with noiseless QAOA, GW, and Iceberg QAOA with four syndrome measurements, as well as unencoded QAOA emulated on the H2-1 emulator. The noiseless QAOA is able to outperform GW for $\ell \geq 6$ layers. However, both the Iceberg and unencoded QAOA have not yet surpassed GW.
\rev{Although more effective heuristic solvers for the MaxCut problem exist~\cite{dunning2018works, schuetz2022combinatorial}, we employ the GW algorithm as our baseline due to its robust theoretical foundations and its widespread use in benchmarking quantum algorithms~\cite{tate2023warm, PhysRevA.103.042612, Wurtz2021}. This discussion does not aim to claim a quantum advantage over classical algorithms. Instead, it illustrates that QED can reduce the physical fidelity requirements needed to match the performance of the GW algorithm.}

Specifically, at $\ell=10$, the approximation ratio of noiseless QAOA is $0.9810\ldots$, while the average approximation ratio for GW is $0.9554\ldots$. This implies that the logical fidelity of a noisy QAOA must be of at least $\sfrac{0.9554}{0.9810} \simeq 0.974$ to outperform GW, assuming our white noise model approximation. In Fig.~\ref{fig:scaling_beat_GW}B, we vary the scaling factor of the model parameters to determine when this breakeven logical fidelity can be achieved. The results indicate that the Iceberg and unencoded circuits require scaling factors of approximately $0.81$ and $0.60$, respectively. Thus, as hardware technology advances, our results suggest that the Iceberg code enables a breakeven fidelity on small MaxCut problems much earlier than unencoded circuits. 

\begin{figure}[t]
    \centering
    \includegraphics[width = \linewidth]{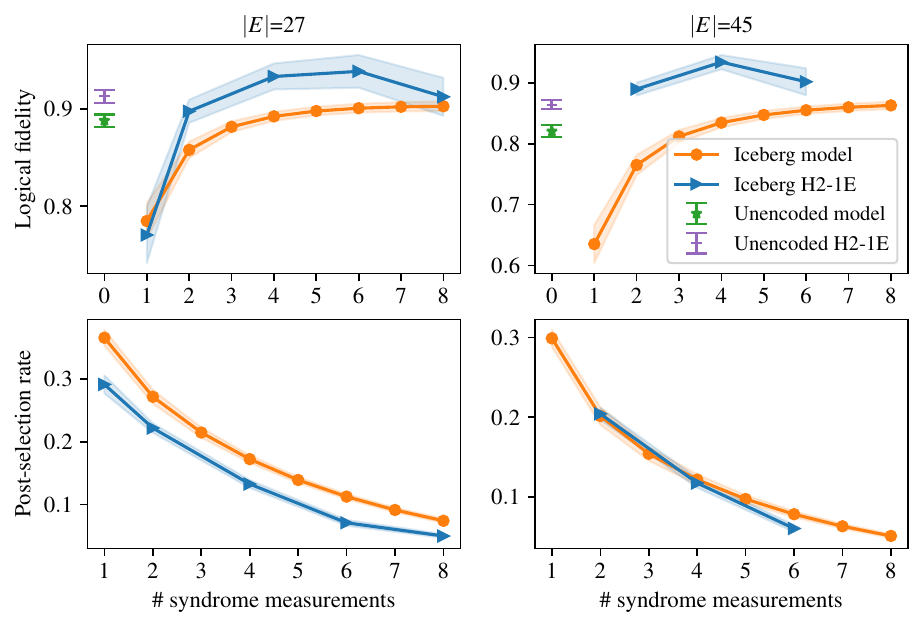}
    \caption{Comparison between \rev{emulated} data and model predictions on random Erd\H{o}s--R\'enyi graphs 
    with different numbers of edges. The prediction of Iceberg logical fidelity is less accurate compared to testing on 3-regular graphs, while the prediction of unencoded logical fidelity and the prediction of post-selection rate remain accurate. 
    } 
    \label{fig:synd_test_discardrate}
\end{figure}
\subsubsection{Model generalization beyond $3$-regular graphs}
So far, we have used 3-regular graphs to fit the performance model and analyzed the model's performance on extrapolated 3-regular graphs. To test the generalization of the model, we validate it on random Erd\H{o}s--R\'enyi graphs, which have different topologies compared to 3-regular graphs. We fix the number of nodes at $k=18$,
set the number of edges as $\lvert E \rvert \in \{27, 45\}$ with all edges generated randomly, and select one graph for each of the two sizes.

We present the comparison between \rev{emulated} results and model predictions in Fig.~\ref{fig:synd_test_discardrate}. The model predictions for the logical fidelity of the Iceberg code circuits are less accurate, specially with the densest graph of $|E|=45$ edges. The model predictions for the unencoded logical fidelity and the Iceberg code post-selection rates are comparably more accurate. This indicates that the fitted model works well for problems with similar topologies, but highlights the limitation of the model's generalization to different problems. We suspect that the worse model performance on these graphs is caused by the different amount of unaccounted noise accumulated in the comparably deeper circuits for these graphs.

\section{Discussion}

We demonstrate that the post-selected samples with the Iceberg code present a higher approximation ratio than the samples from unencoded circuits. This allows us to study the performance of QAOA in an effective noise regime closer to the noiseless computation (at least for circuits on the beneficial side of the breakeven frontier). 
We can see the current breakeven frontier of the Iceberg QAOA on 3-regular graphs in Fig.~\ref{fig:panel-3}A. For example, for $6$ syndrome measurements, the breakeven frontier is up to $k = 20$ logical qubits for $\ell \in [4,12]$ of QAOA layers. However, there are multiple opportunities to improve this result and achieve improved performance with Iceberg code as compared to the unencoded circuits for larger qubit counts.

First, other problems may be more amenable to Iceberg code. Specifically, sparse-graph problems (like the MaxCut of 3-regular graphs discussed in this paper) are not particularly beneficial for the Iceberg code compared to fully-connected Hamiltonians like the Sherrington-Kirkpatrick model \rev{(see an example in Appendix~\ref{app:additional_experiments})}. This is because the Iceberg code can execute two-qubit logical rotations with no overhead, whereas one-qubit logical rotations require a noisier two-qubit physical gate. 
Iceberg QAOA could be more advantageous for dense graph problems where the number of two-qubit $Z^{\otimes2}$ terms in the problem Hamiltonian is much larger than the number of single-qubit $X$ terms in the mixing Hamiltonian.

Second, the compilation of Iceberg circuits could be further optimized. We speculate that the high deviations of the error rates observed between the emulator and fitted error rates in Table~\ref{tab:fitted_para} could be explained by a larger amount of memory noise accumulated in the highly sequential syndrome measurement blocks and the QAOA mixing layer than in the optimized unencoded circuits. Currently, we are using pytket~\cite{sivarajah2020t} to compile the logical gates alone, but a better strategy that jointly compiles the logical gates and the error detection blocks of the Iceberg code could potentially improve hardware performance while reducing execution time. 

Though performance extrapolation indicates promising results with improved hardware fidelity, we observe that with fixed model parameters, increasing the number of syndrome measurements marginally diminishes the performance gains in the extrapolation heatmaps. Additionally, the overhead of post-selecting samples grows rapidly. This observation is consistent with the \rev{emulated} results on the H2-1 emulator, indicating that the protective power of the Iceberg error detection code is limited. This reinforces the need for quantum error correction to achieve error rates low enough to run large-scale circuits.

Benchmarking the performance of a quantum algorithm beyond the simulatable region is inherently challenging, particularly when the algorithm is protected by QED or QEC. To address this, we develop a pipeline that calibrates a reliable performance model and extrapolates it to large-scale circuits. We believe this pipeline could inspire further studies on the performance of fault-tolerant implementations of other algorithms.

\section{Methods}
This section provides further details on the \rev{circuit} realization, presents the model construction, and details the fitting to \rev{emulated} data.

\subsection{Location of syndrome measurements}
As depicted in Fig.\ref{fig:panel-1}A , in this work we place syndrome measurements evenly spaced in the circuit so that every block of logical gates has roughly the same number of logical gates.

\begin{figure}[t]
    \centering
    \includegraphics[width = 0.9\linewidth]
    {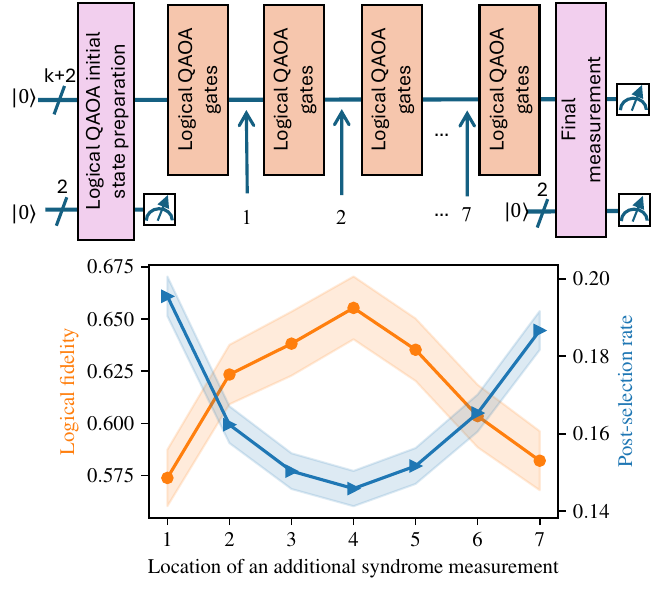}
    \caption{Top: Circuits with different locations for a single syndrome measurement, labeled 1-7. Bottom: \rev{Emulated} performance of the different circuits. The circuit with the syndrome measurement in the middle (labeled 4) detects the most errors and achieves the best logical fidelity. All data include the final measurement and fault-tolerant initialization. Without any syndrome measurement, the logical fidelity is $0.5128 \pm 0.0115$ and the post-selection rate is $0.2762 \pm 0.0058$. With one additional syndrome measurement, all circuits outperform the one without the syndrome measurement. 
    } 
    \label{fig:syndloc_withfinal_single}
\end{figure}

In Fig.~\ref{fig:syndloc_withfinal_single}, we provide evidence supporting this strategy. The $k=16, \ell=11$ logical QAOA circuit is evenly partitioned into eight blocks of logical gates, and we introduce a single syndrome measurement in the seven intermediate positions between them. By comparing the logical fidelity obtained from these seven circuits, we find that the circuit with the syndrome measurement inserted in the middle (labeled 4) presents the highest fidelity at the cost of the smallest post-selection rate. 
\subsection{Performance model} \label{ssec:model}
This section introduces the detailed model to predict the performance of the unencoded and Iceberg code circuits. To build the model outputs efficiently, we consider the following noise channels:
\begin{itemize}
    \item Uniform two-qubit depolarizing channel with error rate $p_\ell$ after every two-qubit logical gate of the unencoded circuit: insert a random Pauli error from the set $\mathcal{P}^{\otimes 2}\setminus \{I^{\otimes 2}\}$ on the gate support.
    \item Uniform two-qubit depolarizing channel with error rate $p_{cx}$ after every two-qubit $\cxgate$ gate in the error detection blocks of the Iceberg code circuit: insert a random Pauli error from the set $\mathcal{P}^{\otimes 2}\setminus \{I^{\otimes 2}\}$ on the gate support.
    \item A noise channel with error rate $p_c$ that introduces a random error that commutes with both stabilizers after every logical gate of the Iceberg code: insert a random Pauli error from the set $\{X^{\otimes 2}, Y^{\otimes 2}, Z^{\otimes 2}\}$ on the gate support.
    \item A noise channel with error rate $p_a$ that introduces a random error that anti-commutes with a stabilizer after every logical gate of the Iceberg code: insert a random Pauli error from the set $\mathcal{P}^{\otimes 2}\setminus \{I^{\otimes 2}, X^{\otimes 2}, Y^{\otimes 2}, Z^{\otimes 2}\}$ on the gate support.
\end{itemize}
Overall, we have one error parameter $p_\ell$ for unencoded circuits, and three error parameters, $p_{cx}$, $p_c$, and $p_a$ for Iceberg circuits. The probabilities of no error $P_0$, one error $P_1$, and two or more errors $P_2$ in a set of $g$ gates with error rate $p$ are respectively defined as
\begin{eqnarray}
    P_0(p,g) &&= (1-p)^g, \label{eq:prob0}\\
    P_1(p,g) &&= g p (1-p)^{g-1}, \label{eq:prob1}\\
    P_2(p,g) &&= 1 - P_0(p,g) - P_1(p,g). \label{eq:prob2}
\end{eqnarray}

We additionally make two simplifications to construct the model outputs efficiently. The first one states that 
\begin{simplification}\label{simp:1}
    Every undetected error produces a logical error, unless the fault-tolerance of the Iceberg code prevents it.
\end{simplification}
For unencoded circuits, since no error is detectable, every error produces a logical error by virtue of this simplification, so we model the \textit{unencoded logical fidelity} as the probability of absolutely no error:
\begin{equation}\label{eq:unencoded_lf}
    \mathcal{F}_\mathrm{une} = P_0(p_\ell,g_2).
\end{equation}
From the fault-tolerant design, every single error in the error detection blocks of the Iceberg code is either detectable or acts trivially, so none of these single errors contribute to the logical infidelity of the Iceberg code. However, two or more errors can produce undetectable errors, that, by virtue of this simplification, produce a logical error. The usually implicit justification of this simplification is that the number of errors that act trivially in the circuit is exponentially smaller in the circuit size than the number of undetectable errors that act non-trivially at the logical level.  

\subsubsection{Model for the Iceberg code} \label{sssec:model}
The model for the Iceberg code incorporates the effects of error detection and fault-tolerance. To construct the model efficiently, we divide the circuit into blocks of initialization, logical gates, syndrome measurement and final measurement, respectively, as depicted in Fig.~\ref{fig:panel-1}. At every block, we consider the probabilities of errors causing
\begin{enumerate}
    \item a \textit{harmless error} $H$ if they excite no ancilla and no stabilizer and act trivially on the state, like a $Z$ error before the measurement of an ancilla,
    \item a \textit{logical error} $L$ that excites no ancilla and no stabilizer but acts non-trivially on the state, like a two-qubit $X^{\otimes2}$ error after a logical gate,
    \item an \textit{exciting error} $E$ that excites a stabilizer but not an ancilla, propagating such an error to the next block without an immediate discard, like a single-qubit $X$ error on a code qubit inserted by the last $\cnotgate$ that acts on that qubit in a syndrome measurement, 
    \item or a \textit{discarding error} $D$ that excites an ancilla, causing an immediate discard, like a single-qubit $X$ error before the measurement of an ancilla. 
\end{enumerate}
In the absence of noise, the state remains in the $+1$ eigenspace of the stabilisers and ancillas. An error that excites some of them brings the state to their $-1$ eigenspace, and makes the error detectable. We say that exciting and discarding errors are both detectable errors, while harmless and logical errors are undetectable. The sum of the four probabilities adds up to 1 at every block. 

\begin{table}[b]
\begin{tabular}{r|m{3em}m{3em}m{3em}m{3em}m{3em}}\hline
\# errors          & 0   & $1$            & $\geq1$        & $\geq1$        & $\geq2$        \\ 
excited ancilla    & no  & no             & no             & yes            & no             \\ 
excite stabilizers & no  & no             & yes            & any            & no             \\
fraction of errors & 1   & $\sfrac{1}{8}$ & $\sfrac{4}{8}$ & $\sfrac{3}{8}$ & $\sfrac{1}{8}$ \\
contribute to      & $H$ & $H$            & $D$            & $E$            & $L$            \\ \hline
\multirowcell{4}{Initialize\\the circuit\\probabilities as} 
& \multicolumn{5}{l}{$H\leftarrow P_0(p_{cx}, n+3) + \sfrac{1}{8}P_1(p_{cx}, n+3)$}             \\
& \multicolumn{5}{l}{$L\leftarrow \sfrac{1}{8}P_2(p_{cx}, n+3)$}                                \\
& \multicolumn{5}{l}{$E\leftarrow \sfrac{3}{8}P_1(p_{cx}, n+3) + \sfrac{3}{8}P_2(p_{cx}, n+3)$} \\        
& \multicolumn{5}{l}{$D\leftarrow \sfrac{1}{2}P_1(p_{cx}, n+3) + \sfrac{1}{2}P_2(p_{cx}, n+3)$} \\   
\hline      
\end{tabular} \caption{Classification of errors in the initialization block.} \label{tab:ini}
\end{table}

\begin{table*}[t]
\centering
\begin{tabular}{r|m{2.5em}m{2.5em}m{2.5em}m{2.5em}m{2.5em}m{2.5em}m{2.5em}m{2.5em}m{2.5em}m{2.5em}m{2.5em}m{2.5em}m{2.5em}m{2.5em}} \hline
input error              & $H$ & $H$     & $H$     & $H$            & $H$            & $L$     & $L$     & $L$            & $L$            & $E$     & $E$            & $E$            & $E$             & $E$             \\ 
\# anti-commuting errors & 0   & 0       & 1       & $\geq2$        & $\geq2$        & 0       & 1       & $\geq2$        & $\geq2$        & 0       & 1              & 1              & $\geq2$         & $\geq2$         \\
\# commuting errors      & 0   & $\geq1$ & $\geq0$ & $\geq0$        & $\geq0$        & $\geq0$ & $\geq0$ & $\geq0$        & $\geq0$        & $\geq0$ & $\geq0$        & $\geq0$        & $\geq0$         & $\geq0$         \\
excited stabilizers      & no  & no      & yes     & no             & yes            & no      & yes     & no             & yes            & yes     & no             & yes            & no              & yes             \\
fraction of errors       & 1   & 1       & 1       & $\sfrac{1}{4}$ & $\sfrac{3}{4}$ & 1       & 1       & $\sfrac{1}{4}$ & $\sfrac{3}{4}$ & 1       & $\sfrac{3}{9}$ & $\sfrac{6}{9}$ & $\sfrac{3}{12}$ & $\sfrac{9}{12}$ \\
contribute to            & $H$ & $L$     & $E$     & $L$            & $E$            & $L$     & $E$     & $L$            & $E$            & $E$     & $L$            & $E$            & $L$             & $E$             \\ \hline
\multirowcell{4}{Update rules\\to add this block} 
 & \multicolumn{14}{l}{\hspace{2em}$H\leftarrow HP_0(p_c, g)P_0(p_a, g)$}                                                         \\
 & \multicolumn{14}{l}{\hspace{2em}$L\leftarrow H(1-P_0(p_c, g))P_0(p_a, g) + LP_0(p_a, g) + \sfrac{1}{3}EP_1(p_a, g) + \sfrac{1}{4}(H + L + E)P_2(p_a, g)$}    \\
 & \multicolumn{14}{l}{\hspace{2em}$E\leftarrow  EP_0(p_a, g) + (H+L+\sfrac{2}{3}E)P_1(p_a, g) + \sfrac{3}{4}(H+L+E)P_2(p_a, g)$} \\
 & \multicolumn{14}{l}{\hspace{2em}$D\leftarrow D$} \\ 
 \hline
\end{tabular}
\caption{Classification of errors in the block of logical gates}\label{tab:log}
\end{table*}

These probabilities are initialized from the initialization block, and iteratively updated for every block of logical gates, syndrome measurement, and the final measurement. For example, when adding a block of $g$ logical gates that can suffer internal errors, the input probability of harmless errors $H$ from previous blocks updates to the joint probability $H \leftarrow HP_0(p_c, g)P_0(p_a, g)$ of an input harmless error and no internal errors. At the end of this iterative process, we obtain analytical functions for the probabilities of the entire circuit, from which we can compute the model outputs for the Iceberg code. 

To construct the model outputs efficiently we make the following additional simplification:
\begin{simplification}\label{simp:2}
    Errors are evenly distributed across all possible excitation events. That is, for every number $\mu=1$ and $\mu\geq2$ of errors in a block with $m\geq0$ ancillas, the probability of all excitation events is the same and equal to $\sfrac{1}{2^{m+2}}$. The excitation events are the $2^{m+2}$ possible ways to excite (or not) the $m$ ancilla(s) and the two stabilizers. 
\end{simplification}
For example, in the syndrome measurement block, there are 16 possible excitation events, depending on which of the two stabilizers and the two ancillas are excited (or not) by errors. Therefore, we assume that $\sfrac{1}{16}$ of the errors excite no ancilla and no stabilizer, $\sfrac{12}{16}$ of them excite an ancilla, and $\sfrac{3}{16}$ excite a stabilizer without exciting the ancillas. This simplification allows the error detection and fault-tolerance properties of the Iceberg code to be incorporated in a very natural way. Appendix~\ref{app:model} show the deviations between the model predictions and the exact fractions computed for small Iceberg code instances.  
\\

\textit{Initialization block.} The top part of Table~\ref{tab:ini} presents the contribution of every possible error in this block. For each number $\mu=1$ and $\mu\geq2$ of errors indicated in the first row (columns 2, 3, 4 and columns 3, 4, 5, respectively), the table distinguishes the possible excitations these errors can cause on the ancilla (second row) and stabilizers (third row). Using Simplification~\ref{simp:2} we compute the fraction of errors that contribute to each of the possible events and present them in the fourth row. The fifth row shows how this classification incorporates the fault-tolerance of the initialization block. Every error is either harmless or detectable, while two or more errors that pass undetected (last column) produce a logical error by virtue of simplification~\ref{simp:1}. Aggregating the contributions provides the initialization of the circuit probabilities presented in the last rows.

\textit{Block of logical gates}. To add a block of logical gates, we need to update the circuit probabilities from the combination of the input errors from the previous blocks with the internal errors of this block. Since no discard is possible at the block of logical gates, the discard probability $D$ is preserved. The top part of Table~\ref{tab:log} presents the classification of input, internal anti-commuting errors, and internal commuting errors. The fraction of errors is calculated again using Simplification~\ref{simp:2}. For undetectable input errors, two or more internal anti-commuting errors cause 4 possible excitation events depending on which of the two stabilizers are excited or not. In contrast, input exciting errors and single anti-commuting errors each create 3 possible stabilizer excitations: only $S_Z$ excited, only $S_X$ excited, or both excited. Together they create 9 possible excitation events. In 3 of them, the excitations cancel out (11th column), while 6 preserve some excitation (12th column). Similarly, two or more anti-commuting errors create 4 possible excitation events, depending on which of the two stabilizers get excited or not. Thus, here 12 
events are possible when combined with an input exciting error. The bottom part of the table presents the resulting update rules of the circuit probabilities to add a block of $g$ logical gates.

\textit{Syndrome measurement block}. Similar to the block of logical gates, the contributions from every combination of input and internal errors are summarized in Appendix~\ref{app:model}. The main difference is that the circuit is immediately discarded if at least one of the two ancillas is excited. 

\textit{Final measurement block.} To add the final measurement block, note that the stabilizer $S_Z$ is computed in post-process, while $S_X$ is measured by the ancilla. Note also that the second ancilla of the block is a flag qubit which can be excited only by internal errors (not by exciting input errors). Contributions are summarized in Appendix~\ref{app:model}.
\\

Finally, after adding all blocks one by one, we obtain the model output analytical expressions for the circuit probabilities $H$, $L$, $D$ (and the trivial $E=0$). The \textit{post-selection rate} is $1-D$ and the \textit{Iceberg code logical fidelity} is 
\begin{equation}
    \mathcal{F}_\mathrm{ice} = H / (1-D).
\end{equation}

We check in Appendix~\ref{app:model} that the first and second order terms at low error rates satisfy the expected behaviour
\begin{equation} \label{eq:leading}
    1 - \mathcal{F}_\mathrm{ice} = (g_1+g_2) p_c + O\left(p_{cx}^2 + p_a^2 + p_{cx}p_a + p_c^2\right).
\end{equation}
This numerically confirms the partially fault-tolerant nature of the Iceberg code: the only single errors capable of causing a logical error are commuting errors in the logical gates, but no single error from the other two noise sources can cause a logical error.

\subsection{Approximation ratio and logical fidelity}  \label{ssec:white_noise_model}
To relate the logical fidelity and the noisy approximation ratio, we consider that the circuit noise takes the form of a global white noise channel~\cite{nielsen2010quantum}
\begin{equation}
    \bm{\rho} = \mathcal{F}\ket{\bm{\psi}}\bra{\bm{\psi}} + (1-\mathcal{F})2^{-k}I^{\otimes k}.
\end{equation}
We can then estimate the noisy approximation ratio from the modeled logical fidelity $\mathcal{F} \in \{\mathcal{F}_\mathrm{une}, \mathcal{F}_\mathrm{ice}\}$ as
\begin{eqnarray} \label{eq:noisy_ar}
    \alpha(\bm{\rho}) = \frac{|E| -  \mathcal{F} \braket{\bm{\psi}|\mathcal{H}|\bm{\psi}} }{2f_{\rm max}}.
\end{eqnarray}
This is the approximation ratio estimated from the performance model that we report in this work.

Moreover, we sometimes use the white noise channel to estimate the logical fidelity from the samples obtained by running the unencoded circuits and the Iceberg code circuits on hardware or the emulator. In this scenario, we have access to the experimental approximation ratio, or equivalently, to the average energies $\braket{\mathcal{H}}_c$ for every QAOA circuit $c \in C_{\rm une}\cup C_{\rm ice}$ considered in this work, unencoded or protected by the Iceberg code. The estimated logical fidelity is the one that reproduces such average energy under the white noise channel:
\begin{equation} \label{eq:L_fromdata}
    \mathcal{F}_c = \frac{\braket{\mathcal{H}}_c}{\braket{\bm{\psi}|\mathcal{H}|\bm{\psi}}}.
\end{equation}
When we have access to the experimental implementation of the QAOA circuits on hardware or the emulator, this is the estimated logical fidelity we report in this work.

Given a graph with edges $E_c$ solved by the QAOA circuit $c$, we additionally consider the set of ratios
\begin{equation}\label{eq:edgewise_fidelity}
    F_c = \left\lbrace \frac{\braket{Z_i Z_j}_c}{\braket{\bm{\psi}|Z_i Z_j|\bm{\psi}}}: \:  (i,j)\in E_c \right\rbrace
\end{equation}
obtained from the experimental expected values $\braket{Z_i Z_j}_c $. Note that in general, these experimental ratios 
are not expected to take the same value for all edges, but under the white noise channel, they all equal the channel fidelity. 

To quantify the deviation between the \rev{emulated} data and the white noise simplification, we consider the normalized distance between the experimental logical fidelity and the ratios as
\begin{eqnarray}
    d\left(F_c, \mathcal{F}_c\right) &&= \frac{1}{\lvert E_c \rvert} \sqrt{l\left(F_c, \mathcal{F}_c\right)}, \rm{with} \label{eq:lf_distance} \\
    l\left(F_c, \mathcal{F}_c\right) &&= \sum_{\mathcal{F}_{ij} \in F_c}{\left(\mathcal{F}_{ij} - \mathcal{F}_c\right)}^2.
\end{eqnarray}

\subsection{Model fitting}

For the unencoded circuits, we use the least squares method as a loss function to minimize the residual between the model logical fidelity $\mathcal{F}_{\rm une}$ and the experimental ratios obtained from the unencoded circuits. For every QAOA instance $c$, we compute the set of ratios $F_{c, \rm une}$ and then aggregate them all to the loss function:
\begin{equation}
   l_\mathrm{une}(\mathcal{F}_{\rm une}) = \sum_{c\in C_{\rm une}} l(F_c, \mathcal{F}_{\rm une}).
\label{eq:fit_lossfunction_une}
\end{equation}
Still leveraging the least squares method to fit the Iceberg code model to the \rev{emulated} data, we minimize the residuals of both the logical fidelity and discard rate. Given a QAOA instance $c \in C_{\rm ice}$ with experimental discard rate $D_c$, we define the loss function for multiple QAOA instances as
\begin{equation}
   l_\mathrm{ice}(\mathcal{F}_{\rm ice}) = \sum_{c\in C_{\rm ice}} \frac{1}{\lvert F_c\rvert}l(F_c, \mathcal{F}_{\rm ice}) + {\left(D_c - D\right)}^2.
\label{eq:fit_lossfunction_ice}
\end{equation}

\section*{Data availability}
Data for reproducing all figures and tables is available at is available at \href{https://zenodo.org/records/15231805}{https://zenodo.org/records/15231805}.

\section*{Code availability}
The code for numerical simulations is available upon reasonable request.

\section*{Acknowledgments} 
We would like to thank the experimental scientists in the Quantinuum team that design and maintain the H-series devices. We extend our gratitude to Drs. Matthew DeCross, Ciaran Ryan-Anderson and Selwyn Simsek for valuable discussions about the modeling of memory errors in these devices and Drs. David Hayes, Changhao Li, Pradeep Niroula, and Shree Hari Sureshbabu for the preliminary statistical analysis of the numerical experiments. We appreciate the comments and suggestions of Drs. Chris N. Self, Michael Perlin and Sivaprasad Omanakuttan on the manuscript. We thank the technical staff at JPMorganChase's Global Technology Applied Research Center for their support and helpful discussions. 
This paper was prepared for informational purposes with contributions from the Global Technology Applied Research center of JPMorgan Chase \& Co. This paper is not a product of the Research Department of JPMorgan Chase \& Co. or its affiliates. Neither JPMorgan Chase \& Co. nor any of its affiliates makes any explicit or implied representation or warranty and none of them accept any liability in connection with this paper, including, without limitation, with respect to the completeness, accuracy, or reliability of the information contained herein and the potential legal, compliance, tax, or accounting effects thereof. This document is not intended as investment research or investment advice, or as a recommendation, offer, or solicitation for the purchase or sale of any security, financial instrument, financial product or service, or to be used in any way for evaluating the merits of participating in any transaction.

\section*{Author Contributions}
Z.H. developed emulation and model fitting pipeline. D.A. developed the performance model and conducted the hardware experiments. R.S. contributed to the data analysis and experimental design. All authors participated in technical discussions and contributed to the writing of the manuscript.

\section*{Competing Interests}
The authors declare no competing financial or non-financial interests.

\bibliography{main}%

\newpage
\appendix 

\section{Iceberg QAOA circuits details} \label{app:iceberg_circuit}
In this appendix, we depict the error detection blocks of the Iceberg code and briefly describe the strategy taken from~\cite{yamamoto2024demonstrating} to mitigate memory noise.

\begin{figure}[b]
    \centering
    \includegraphics[width = \linewidth]{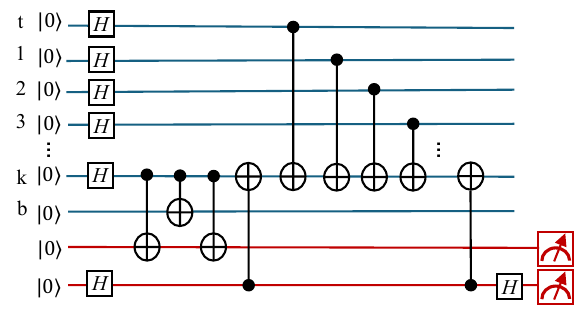}
    \caption{Circuit for fault-tolerant initialization of the logical state $\ket{\bar{+}}^{\otimes k}$.} 
    \label{fig:ft_init_plus}
\end{figure}

\begin{figure}[t]
    \centering
    \includegraphics[width = \linewidth]{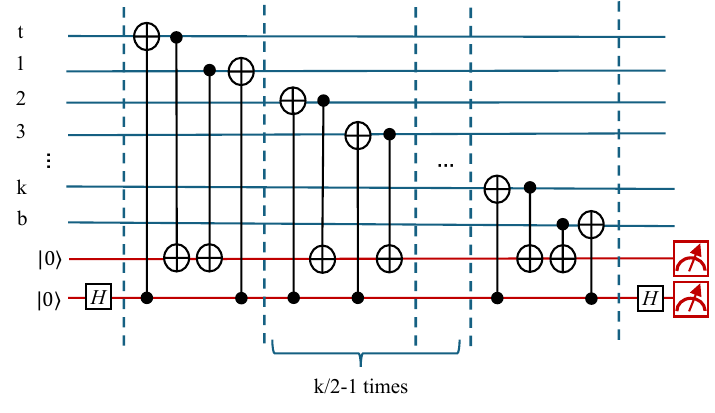}
    \caption{Circuit for fault-tolerant syndrome measurement.} 
    \label{fig:ft_synd_meas}
\end{figure}

\begin{figure}[t]
    \centering
    \includegraphics[width = \linewidth]{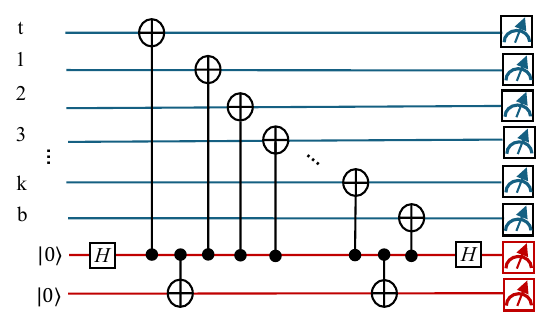}
    \caption{Circuit for fault-tolerant final measurement.} 
    \label{fig:ft_meas}
\end{figure}

\textit{Error detection blocks of the Iceberg code.}
The initialization in Fig.~\ref{fig:ft_init_plus}, taken from~\cite{chao2018quantum}, prepares all logical qubits in the initial logical state $\ket{\bar{+}}^{\otimes k}$. If any of the two ancilla measurements (indicated in red) is $-1$, an error is detected and the circuit is discarded. The syndrome measurement in Fig.~\ref{fig:ft_synd_meas} and the final measurement in Fig.~\ref{fig:ft_meas} were taken from~\cite{self2024protecting}. Both circuits measure the two Iceberg code stabilizers in order to detect errors, indicated again by excited ancilla measurements. The measurement of the $S_Z$ stabilizer is performed in the post-process of the $n$ destructive measurement outcomes of the code qubits, where its value is the parity of the outputs. If no errors are detected, the logical outcomes are extracted as well from pairwise parities. 
These circuits are fault-tolerant, so that every single error from their components (\cnotgate, single-qubit gates, qubit initialization, qubit measurement, or idle qubits) either acts trivially without causing a logical error, or is detectable. Consequently, at least two errors are necessary to cause a logical error. 

\textit{Dynamic decoupling to suppress memory error.}
As demonstrated in~\cite{yamamoto2024demonstrating}, physical memory errors significantly impact both logical fidelity and post-selection rate. On the Quantinuum H2-1 emulator, memory errors are primarily modeled as coherent single-qubit rotations $\exp(-i \epsilon Z)$ of a small angle $\epsilon \ll 1$. We follow the same strategy to mitigate these errors: we apply the stabilizer $S_X$, i.e., a layer of $X$ gates acting on all physical qubits, between all QAOA layers. The stabilizer acts trivially on the logical state, but is expected to mitigate memory errors. 

\section{Experimental details} \label{app:data}
This appendix discusses the gap observed between the hardware and emulator performance, provides some detail about the filtering of data for the model fitting, presents a detailed validation of the model fitting and the white noise simplification, and comments on an early attempt to employ a simpler model. 

\begin{figure}[t]
    \centering
    \includegraphics[width = \linewidth]{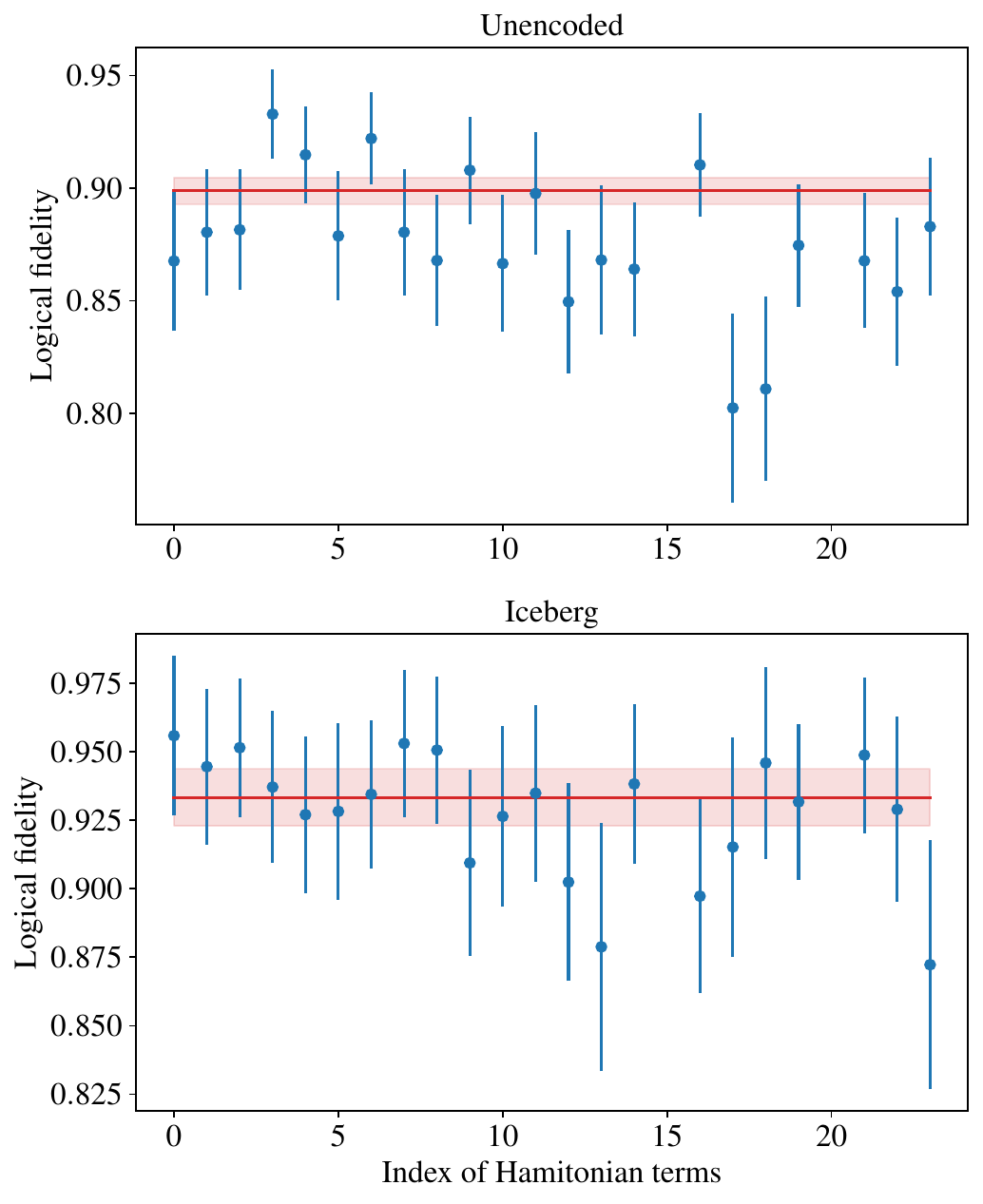}
    \caption{Validation of the white noise assumption on $k=16$, $\ell=10$ unencoded and Iceberg QAOA circuits. The Iceberg circuit has $4$ syndrome measurements. The blue scatters show the energy ratios of each Hamiltonian term (excluding a few outliers) obtained from Eq.~\eqref{eq:edgewise_fidelity}.
    The red line represents the overall logical fidelity obtained from Eq.~\eqref{eq:L_fromdata}. The error bars and shaded regions represent the standard errors. %
    } 
    \label{fig:check_lf_assumption}
\end{figure}

\begin{figure}[t]
    \centering
    \includegraphics[width =\linewidth]{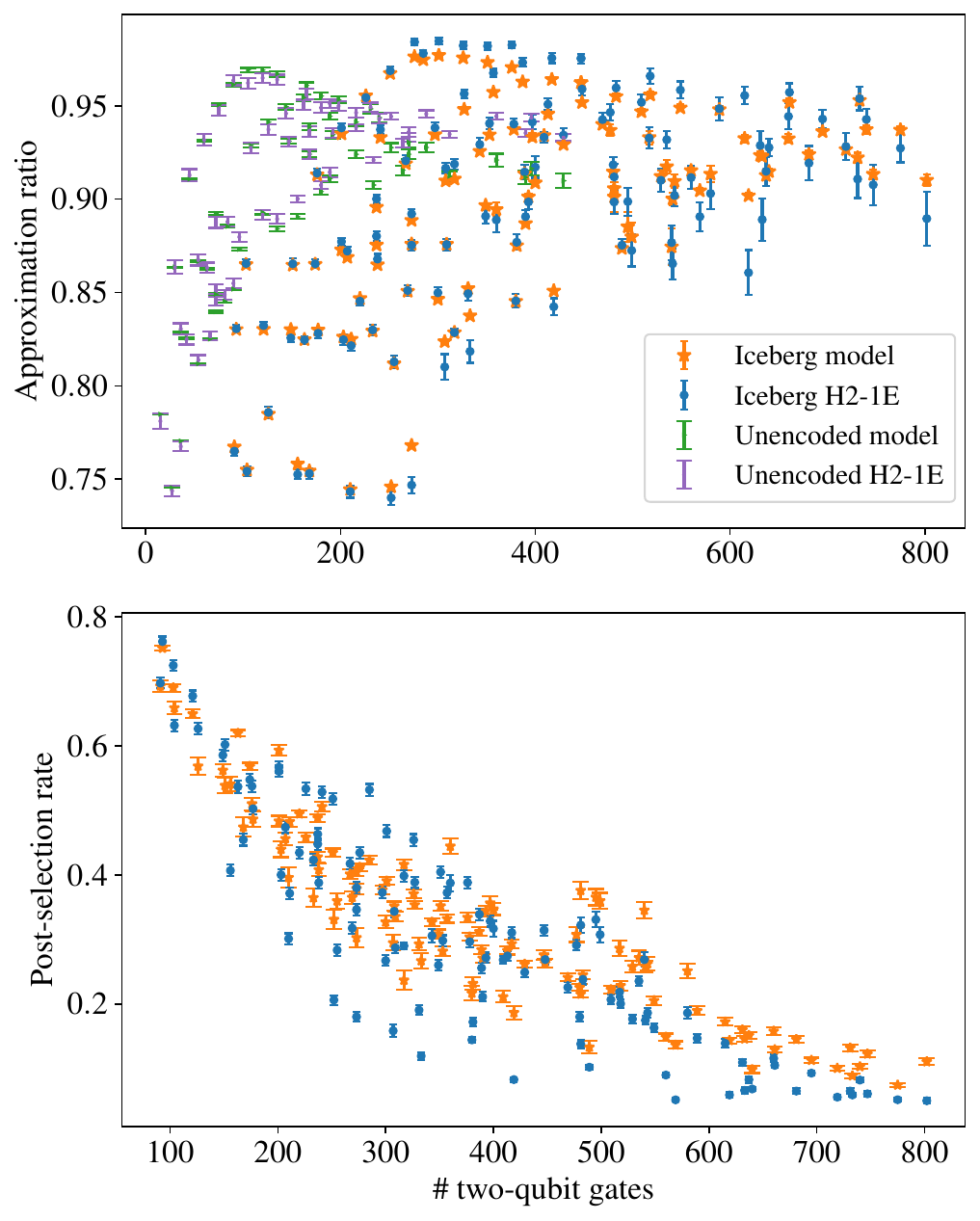}
    \caption{Model predictions of the approximation ratio, calculated by Eq.~\eqref{eq:noisy_ar}, and post-selection rates on all collected \rev{emulated} data. A portion of these data points are used as training data. The scatter plots are ordered by the number of two-qubit gates in the compiled circuits. The error bars represent the standard errors.
    The average $L_2$ norm of the noisy approximation ratio error for unencoded circuits, the approximation ratio error for Iceberg circuits, and the post-selection rate error for Iceberg circuits are 1.2e-3, 9.0e-4, and 5.5e-3, respectively.}
    \label{fig:fitting_result}
\end{figure}

\textit{Gap between hardware and emulator.}
In Table~\ref{tab:gap_hardware_emulator}, we compare the performance of H2-1 device and emulator experiments on both Iceberg and unencoded circuits. These QAOA circuits are fixed with $\ell=10$ and have varying number $k$ of logical qubits. The Iceberg circuits use four syndrome measurements. The differences in logical fidelity between H2-1 device and emulator are small for unencoded circuits but become increasingly significant for Iceberg code circuits as $k$ increases, where the hardware moderately outperforms the emulator. The emulator consistently estimates larger values of the post-selection ratio.
The H2-1 hardware experiments in Fig.~\ref{fig:panel-1}C show that for up to $k=20$ and $\ell=10$ with four syndrome measurements, the Iceberg circuits outperform the unencoded circuits. However, this performance is beyond the frontier observed in the H2-1 emulator experiments and the model predictions. This discrepancy is likely due to the fidelity gap between the H2-1 device and emulator when handling relatively large-sized circuits that have both large qubit count and circuit depth. 

\textit{Filtering fitting data.}
We conduct two levels of filtering on the fitting data. At the first level, we apply two criteria to filter QAOA circuits: the number of two-qubit gates and the uncertainty of the experimental logical fidelity as defined in Eq.~\eqref{eq:L_fromdata}. We use relatively large-sized circuits to fit the model, as we are more interested in capturing and predicting their behavior, such as compiled Iceberg circuits with $\geq 150$ two-qubit gates and compiled unencoded circuits with $\geq 200$ two-qubit gates. Additionally, we aim to use high-quality logical fidelity $\mathcal{F_\mathrm{une}}$ and $\mathcal{F_\mathrm{ice}}$ in the loss functions~\eqref{eq:fit_lossfunction_une} and~\eqref{eq:fit_lossfunction_ice}. Therefore, we set a threshold for the ratio of the standard error to the mean of the logical fidelity, which is $0.01$ for unencoded circuits and $0.012$ for Iceberg circuits. This threshold is particularly important for Iceberg circuits, as the number of experimental post-selected samples becomes limited for large-sized circuits, leading to a larger standard error.
At the second level, given the selected QAOA circuits, we filter out outlier Hamiltonian terms whose experimental logical fidelity $\mathcal{F}_{ij}$ is either greater than the noiseless fidelity or less than half of the noiseless fidelity. This effectively filters out Hamiltonian terms whose energy does not conform to the white-noise approximation.
As mentioned above, we filter some data before fitting the model. Here, in Fig.~\ref{fig:fitting_result}, we present a comprehensive comparison of the model predictions and \rev{emulated} data on all available data.

\textit{Validation of white noise assumption.}
To validate the white noise assumption in Sec.~\ref{ssec:white_noise_model}, we plot the ratios in Eq.~\eqref{eq:edgewise_fidelity} in Fig.\ref{fig:check_lf_assumption} against the white noise fidelity of Eq.~\eqref{eq:L_fromdata}. If the \rev{emulated} data followed a perfect white noise distribution, all ratios would coincide with the white noise fidelity, but this is clearly not the case. However, the normalized distances between them are as small as 8.1e-3 $\pm$ 1.7e-3 and 5.0e-3 $\pm$ 1.9e-3 for the unencoded and Iceberg circuits, respectively, making the white noise model a reasonable simplification for the purpose of this work.

\textit{Comparison with previous model.}
For Iceberg circuits, leveraging the model in Sec.~\ref{sssec:model}, we can accurately predict both the logical fidelity and post-selection rates using a single set of parameters, as shown in Table~\ref{tab:fitted_para}. We also attempted to fit the performance model reported in~\cite{proctor2022measuring}. While this model can fit the logical fidelity and post-selection rate separately with decent accuracy, it fails to fit both quantities simultaneously with one set of parameters.

\rev{\textit{Runtime Estimation.}
It is also intriguing to examine the runtime of the Iceberg QAOA circuit compared to an unencoded circuit. Since samples may be discarded during the execution of the Iceberg program if fault-tolerant gadgets detect errors, let $t_{ps}$ represent the execution time of a circuit where all sample discarding is deferred to the end of the program. We introduce a factor $\lambda$ to quantify the runtime reduction of a program with discarding, denoted as $t_{ice}$:
\begin{align}
    t_{ice} & = \lambda * t_{ps} \\
    \lambda & = \sum_{i}^{g} d_i t_i / t_{ps}
\end{align}
where $g$ is the total number of FT gadets (including FT initialization, syndrome measurement, and FT final measurement), $d_i$ is the ratio of samples that end execution at the $i$-th fault-tolerant (FT) gadget with $\sum_i^g d_i = 1$, and $t_i$ is the runtime up to the $i$-th gadget. As the last gadget is the FT final measurement, we have $t_{g} = t_{ps}$.}

\rev{
While the actual hardware execution time of programs is typically not accessible to users, we estimate the execution time of the circuits using data from the H2-1 emulator. This emulator reports the duration of a single circuit shot by accounting for both quantum gate operations and classical operations, such as ion movement and cooling. Additionally, by tracking all syndrome measurements during the emulations, we can readily calculate the ratios $d_i$.}

\begin{table*}[h]
\centering
\begin{tabular}{cc|ccccc}
\hline
& $k$ & 16 & 18 & 20 & 22 & 24 \\ \hline
\multirow{5}{*}{Iceberg} 
& Fidelity H2-1 & $0.929 \pm 0.017$ & $0.925 \pm 0.018$ & $\bm{0.901} \pm 0.028$ & $\bm{0.885} \pm 0.020$ & $\bm{0.871} \pm 0.061$ \\ %
& Fidelity H2-1E & $\bm{0.933} \pm 0.010$  &  $\bm{0.945} \pm 0.011$ & $0.86 \pm 0.022$ & $0.835 \pm 0.025$ & $0.803 \pm 0.029$ \\ \cline{2-7}
& Shots H2-1  & 137/1624 & 174/1987 & 55/2000	 & 117/3000 & 17/1158 \\ 
& Shots H2-1E & 489/3000 & 417/3000 & 195/3000 & 182/3000 & 158/3000 \\
& Post-selection H2-1 & $0.084 \pm 0.007$ & $0.088 \pm 0.006$ & $0.028 \pm 0.004$ & $0.039 \pm 0.004$ & $0.015 \pm 0.004$ \\ 
& Post-selection H2-1E & $\bm{0.163} \pm 0.007$ & $\bm{0.139} \pm 0.006$ & $\bm{0.065} \pm 0.005$ & $\bm{0.061} \pm 0.004$ & $\bm{0.053} \pm 0.004$\\
\cline{2-7}
& Gate count & 549 & 615 & 681 & 747 & 813 \\ %
\hline
\hline
\multirow{5}{*}{Unencoded} 
& Fidelity H2-1 & $0.891 \pm 0.008$ & $0.901 \pm 0.006$ & $0.877 \pm 0.008$ & $0.895 \pm 0.008$ & $0.906 \pm 0.007$ \\ %
& Fidelity H2-1E & $0.899 \pm 0.006$ & $0.900 \pm 0.006$ & $0.902 \pm 0.008$ & $0.900 \pm 0.007$ & $0.909 \pm 0.005$ \\ \cline{2-7}
& Shots H2-1 & 500 & 1000 & 500 & 500 & 500 \\ %
& Shots H2-1E & 1000 & 1000 & 1000 & 1000 & 1000 \\ \cline{2-7}
& Gate count & 240 & 270 & 300 & 330 & 360 \\ \hline
\end{tabular}
\caption{The performance of both Iceberg and unencoded QAOA under H2-1 device and its emulator H2-1E. For the Iceberg code, the bar $/$ separates the number of post-selected shots from the total number of submitted shots. 
The values after $\pm$ present standard errors.
}
\label{tab:gap_hardware_emulator}
\end{table*}
\rev{For the five instances listed in Table.~\ref{tab:gap_hardware_emulator} with $k=$16, 18, 20, 22, 24, their runtime results are summarized in Table.~\ref{tab:runtime_res}. The non-monotonic change in the runtime of unencoded circuits $t_{une}$ as $k$ increases is attributed to the sub-optimality of the pytket compilation pass. For these instances, the time per shot of an encoded circuit, $t_{ice}$, is approximately 1.9 to 3 times greater than $t_{une}$. Additionally, more total shots are required to obtain a sufficient number of post-selected shots. For example, in the emulation experiments of these five instances, we set a total of 3000 shots for encoded circuits and 500 shots for unencoded ones in emulations, resulting in an overall slowdown of 12 to 18 times for Iceberg QAOA compared to the unencoded version. This demonstrates a trade-off between computational time and improved performance.}

\rev{While the hardware execution runtime of a QED/QEC encoded circuit is a critical consideration for practical deployment, it is important to note that we do not currently claim any speedup for QED circuits. The primary purpose of applying Iceberg encoding is to aid in benchmarking the QAOA algorithm on hardware. We have provided a basic estimation of execution time above, but a more comprehensive study on runtime is deferred to future work.}

\begin{table*}[!h]
\centering
\resizebox{\textwidth}{!}{
\begin{tabular}{c|cccccc}
\hline
Perturbation level $\delta$   & 0 & 0.1   & 0.2   & 0.3   & 0.5   & 1     \\
\hline
Noiseless & 0.988 & 0.969 & 0.924 & 0.855 & 0.659 & 0.544 \\
Encoded H2-1E  & 0.958 $\pm$ 0.005  & 0.955 $\pm$ 0.004 & 0.905 $\pm$ 0.005 & 0.832 $\pm$ 0.006 & 0.659 $\pm$ 0.006 & 0.542 $\pm$ 0.005\\
Unencoded H2-1E & 0.943 $\pm$ 0.003 & 0.930 $\pm$ 0.004 & 0.890 $\pm$ 0.006 & 0.833 $\pm$ 0.006 & 0.649 $\pm$ 0.006 & 0.543 $\pm$ 0.005 \\
\hline
\end{tabular}}
\caption{
\rev{Performance of $\ell=10$ QAOA with perturbed parameters on a $k=16$ graph instance: The values following the $\pm$ symbol represent standard errors.}
}
\label{tab:perturb_qaoa}
\end{table*}
\rev{\section{Additional experiments}} \label{app:additional_experiments}
\rev{\textit{Performance under suboptimal QAOA parameters.}
In this paper, we utilized a fixed set of high-quality QAOA parameters for benchmarking. It is also insightful to examine the performance of the Iceberg code when the parameters are not highly optimized. We conducted an emulation to assess the performance of Iceberg QAOA when parameters are perturbed from their fixed angles, defined as:
\begin{align}
    \bm\Tilde{\gamma} = \bm\gamma + \delta \mathcal{U} \\
    \bm\Tilde{\beta} = \bm\beta + \delta \mathcal{U},
\end{align}
where $\delta$ represents the perturbation level, and $\mathcal{U}$ is a random noise uniformly distributed in $[0,1]$. We selected a $k=16$ MaxCut instance and executed both unencoded QAOA and Iceberg QAOA with $\ell=10$ under parameters with varying levels of perturbation. The approximation ratios for noiseless QAOA, noisy Iceberg QAOA, and noisy unencoded QAOA are presented in Table~\ref{tab:perturb_qaoa}. As the perturbation level $\delta$ increases, both encoded and unencoded QAOA show decreasing approximation ratios, with the advantages of Iceberg encoding diminishing. This suggests that when the QAOA parameters are slightly away from the optimum, we expect that QED will still be helpful in estimating the QAOA energy, thereby benefiting the variational parameter optimization~\cite{RevModPhys.95.045005,robbiati2023real}.}

\begin{figure}[!t]
    \centering
    \includegraphics[width=\columnwidth]{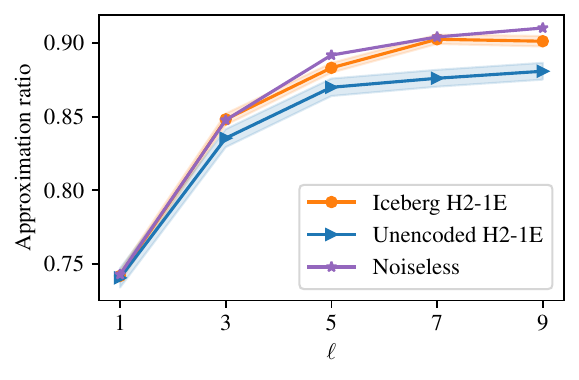}
    \caption{\rev{Emulated performance of Iceberg QAOA and unencoded QAOA on an SK problem with $k=10$. Iceberg QAOA, implemented with $s=4$, outperforms the unencoded one when $\ell \geq 3$, as accumulated noise starts to negatively impact the unencoded QAOA.}}
    \label{fig:sk_result}
\end{figure}

\rev{\textit{Iceberg QAOA on the Sherrington-Kirkpatrick problem.}
This paper primarily demonstrates the performance of Iceberg QAOA on the MaxCut problem, but the approach can be readily extended to other problem families. In addition, we showcase the application of Iceberg QAOA on the Sherrington-Kirkpatrick (SK) model, another widely used benchmark for quantum optimization algorithms. The problem Hamiltonian is given by:
\begin{equation}
    \mathcal{H} = \sum_{i<j} J_{i,j} Z_i Z_j,
\end{equation}
where $J_{i,j}$ follows a standard normal distribution. We employ both Iceberg and unencoded QAOA to solve a $k=10$ SK problem instance using a set of fixed parameters as reported in~\cite{basso2022qaoaskmaxcut}.} 

\rev{As illustrated in Fig.~\ref{fig:sk_result}, when varying the depth $\ell$, Iceberg QAOA with $s=4$ significantly outperforms the unencoded version. At $\ell=1$, their performance is similar, as the circuit is too shallow to manifest the effects of noise.}

\begin{figure}[t]
    \centering
    \includegraphics[width=\columnwidth]{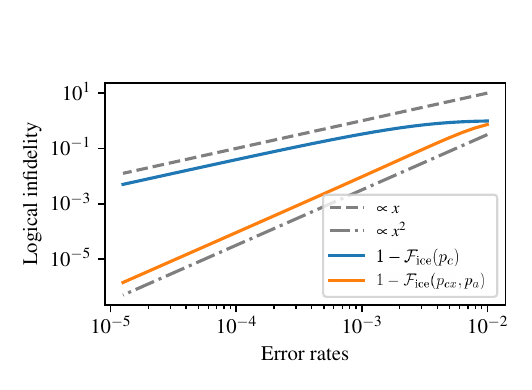}
    \caption{Leading order validation. The Iceberg code infidelity $1-\mathcal{F}_\mathrm{ice}$ is plotted against $p_c$ at $p_{cx}=p_a=0$ in blue, and against $p_{cx}=p_a$ at $p_c=0$ in orange. The top and bottom dashed lines are a linear function $y(x) \propto x$ and a quadratic function $y(x) \propto x^2$ for reference. 
    }
    \label{fig:leading}
\end{figure}
\section{Model details} \label{app:model} 
This appendix numerically justifies Simplification~\ref{simp:2}, provides the derivations introduced in Sec.~\ref{sssec:model} for the syndrome and final measurement blocks, and confirms the leading order stated in Eq.~\eqref{eq:leading} with Fig.~\ref{fig:leading}.

\textit{Justifying Simplification~\ref{simp:2}.}
This simplification can be numerically tested for small number $k$ of logical qubits and number $\mu$ of faulty gates. Table~\ref{tab:simp2} shows a comparison of the simplified model fractions against these numerical values. For the initialization block, the fractions can be read directly from Table~\ref{tab:ini}. For the syndrome and the final measurement blocks, only harmless $H$ input errors are considered, so the model fractions can be read directly from the first columns of Tables~\ref{tab:syn} and~\ref{tab:fin}. The table shows small deviations between the model and the numerical fractions, with the model tending to overestimate the fraction of undetectable errors ($H$ and $L$). However, the deviations reduce quickly as the number $k$ of logical qubits or the number $\mu$ of faulty gates increases. 

\textit{Syndrome measurement block}. Table~\ref{tab:syn} summarises the contributions from input and internal errors. For undetectable input errors, internal errors cause 16 possible excitation events depending on which of the two ancillas and two stabilizers are excited or not. Since input exciting errors produce 3 possible ancilla excitations, the number of combined events grows to 48. If the ancillas excited by the input exciting error and by internal errors coincide, the circuit is not discarded at this block.

\textit{Final measurement block}. Contributions are summarized in Table~\ref{tab:fin}. For input undetectable errors, internal errors produce 8 excitation events depending on which of the two stabilizers and the flag are excited or not. For input exciting errors, the number of excitation events grows to 24. Finally, since no errors can be propagated to the next block, every detectable error at this block produces an immediate discard. We consequently set the update rule $E\leftarrow0$. From the contributions at the top of the table, we obtain the update rules in the bottom part of the table.

\begin{table*}[!h]
\centering
\begin{tabular}{c|ccccc}
\hline
$k$    & 16 & 18 & 20 & 22 & 24 \\
\hline
$\lambda$    & 0.692 & 0.633 & 0.588 & 0.523 & 0.513   \\
$t_{ps} (s)$  & 2.6 & 3.4 & 3.6 & 4.8 & 5.5  \\
$t_{ice} (s)$ & 1.8 & 2.2 & 2.3 & 2.5 & 2.8   \\
\hline
$t_{une} (s)$ & 0.6 & 0.9  &  1.2 & 1.3 & 1.1  \\
\hline
\end{tabular}
\caption{\rev{Estimated runtime of the five instances shown in Table.~\ref{tab:gap_hardware_emulator}}.}
\label{tab:runtime_res}
\end{table*}

\begin{table*}[h]
\centering
\begin{tabular}{c|cccc|cccc|cccc} \hline
Block    & \multicolumn{4}{c|}{Initialization}    & \multicolumn{4}{c|}{Syndrome}  & \multicolumn{4}{c}{Final measurement} \\
Contribution     & $H$     & $L$     & $E$     & $D$     & $H$      & $L$      & $E$      & $D$      & $H$     & $L$     & $E$ & $D$         \\ \hline
Model $\mu=1$    & $0.125$ & $0$     & $0.375$ & $0.5$   & $0.0625$ & $0$      & $0.1875$ & $0.75$   & $0.125$ & $0$     & $0$ & $0.875$     \\
$k=8$, $\mu=1$   & $0.077$ & $0$     & $0.390$ & $0.533$ & $0.0133$ & $0$      & $0.2133$ & $0.7733$ & $0.089$ & $0$     & $0$ & $0.911$     \\
$k=14$, $\mu=1$  & $0.075$ & $0$     & $0.392$ & $0.533$ & $0.0095$ & $0$      & $0.2095$ & $0.7810$ & $0.083$ & $0$     & $0$ & $0.917$     \\
$k=20$, $\mu=1$  & $0.072$ & $0$     & $0.395$ & $0.533$ & $0.0061$ & $0$      & $0.2061$ & $0.7878$ & $0.078$ & $0$     & $0$ & $0.922$     \\
$k=26$, $\mu=1$  & $0.071$ & $0$     & $0.396$ & $0.533$ & $0.0048$ & $0$      & $0.2048$ & $0.7908$ & $0.076$ & $0$     & $0$ & $0.924$     \\ \hline
Model $\mu\geq2$ & $0$     & $0.125$ & $0.375$ & $0.5$   & $0$      & $0.0625$ & $0.1875$ & $0.75$   & $0$     & $0.125$ & $0$ & $0.875$     \\ 
$k=8$, $\mu=2$   & $0.031$ & $0.069$ & $0.293$ & $0.607$ & $0.0055$ & $0.0594$ & $0.1849$ & $0.7502$ & $0.039$ & $0.089$ & $0$ & $0.872$     \\
$k=14$, $\mu=2$  & $0.027$ & $0.081$ & $0.314$ & $0.578$ & $0.0040$ & $0.0615$ & $0.1856$ & $0.7489$ & $0.033$ & $0.095$ & $0$ & $0.872$     \\
$k=20$, $\mu=2$  & $0.026$ & $0.087$ & $0.327$ & $0.560$ & $0.0035$ & $0.0624$ & $0.1859$ & $0.7483$ & $0.030$ & $0.098$ & $0$ & $0.872$     \\
$k=26$, $\mu=2$  & $0.025$ & $0.091$ & $0.335$ & $0.549$ & $0.0032$ & $0.0629$ & $0.1860$ & $0.7479$ & $0.028$ & $0.100$ & $0$ & $0.872$     \\ %
$k=8$, $\mu=3$   & $0.011$ & $0.076$ & $0.260$ & $0.653$ & $0.0008$ & $0.0616$ & $0.1877$ & $0.7499$ & $0.013$ & $0.111$ & $0$ & $0.875$     \\
$k=14$, $\mu=3$  & $0.008$ & $0.089$ & $0.291$ & $0.612$ & $0.0004$ & $0.0620$ & $0.1877$ & $0.75$   & $0.009$ & $0.115$ & $0$ & $0.875$     \\ \hline
    \end{tabular}
    \caption{A comparison of the fractions of errors that contribute to the four block probabilities, as predicted by the model and as calculated exactly for a small number of logical qubits and faulty gates. Only the first significant digits are reported but the values are exact; they do not carry any uncertainty.}
    \label{tab:simp2}
\end{table*}

\begin{table*}[h]
    \centering
\begin{tabular}{r|m{2.5em}m{2.5em}m{2.5em}m{2.5em}m{2.5em}m{2.5em}m{2.5em}m{2.5em}m{2.5em}m{2.5em}m{2.5em}m{2.5em}m{2.5em}m{2.5em}} \hline
input error         & $H$ & $H$             & $H$             & $H$              & $H$             & $L$ & $L$             & $L$             & $L$              & $L$             & $E$ & $E$             & $E$             & $E$              \\ 
\# errors           & 0   & $1$             & $\geq1$         & $\geq1$          & $\geq2$         & 0   & $1$             & $\geq1$         & $\geq1$          & $\geq2$         & 0   & $\geq1$         & $\geq1$         & $\geq1$          \\
excited ancillas    & no  & no              & no              & yes              & no              & no  & no              & no              & yes              & no              & yes & no              & no              & yes              \\
excited stabilizers & no  & no              & yes             & any              & no              & no  & no              & yes             & any              & no              & any & no              & yes             & any              \\
fraction of errors  & 1   & $\sfrac{1}{16}$ & $\sfrac{3}{16}$ & $\sfrac{12}{16}$ & $\sfrac{1}{16}$ & 1   & $\sfrac{1}{16}$ & $\sfrac{3}{16}$ & $\sfrac{12}{16}$ & $\sfrac{1}{16}$ & 1   & $\sfrac{3}{48}$ & $\sfrac{9}{48}$ & $\sfrac{36}{48}$ \\
contribute to       & $H$ & $H$             & $E$             & $D$              & $L$             & $L$ & $L$             & $E$             & $D$              & $L$             & $D$ & $L$             & $E$             & $D$              \\ \hline
\multirowcell{4}{Update rules\\to add this block}
 & \multicolumn{14}{l}{\hspace{5em}$H\leftarrow H\left(P_0(p_{cx}, 2n) + \sfrac{1}{16}P_1(p_{cx}, 2n)\right)$}                               \\
 & \multicolumn{14}{l}{\hspace{5em}$L\leftarrow LP_0(p_{cx}, 2n) + \sfrac{1}{16}(L+E)P_1(p_{cx}, 2n) + \sfrac{1}{16}(H+L+E)P_2(p_{cx}, 2n)$} \\
 & \multicolumn{14}{l}{\hspace{5em}$E\leftarrow  \sfrac{3}{16}(H+L+E)\left(1 - P_0(p_{cx}, 2n)\right)$}                                      \\
 & \multicolumn{14}{l}{\hspace{5em}$D\leftarrow D+ EP_0(p_{cx}, 2n) +\sfrac{3}{4}(H+L+E)\left(1 - P_0(p_{cx}, 2n)\right)$} \\ \hline
\end{tabular}
    \caption{Classification of errors in the syndrome measurement block.}
    \label{tab:syn}
\end{table*}

\begin{table*}[h]
    \centering
\begin{tabular}{r|m{3em}m{3em}m{3em}m{3em}m{3em}m{3em}m{3em}m{3em}m{3em}m{3em}m{3em}m{3em}} \hline
input error        & $H$   & $H$            & $H$            & $H$            & $L$   & $L$            & $L$            & $L$            & $E$ & $E$             & $E$              \\ 
\# errors          & 0     & 1              & $\geq1$        & $\geq2$        & 0     & 1              & $\geq1$        & $\geq2$        & 0   & $\geq1$         & $\geq1$          \\
excitations        & no    & no             & yes            & no             & no    & no             & yes            & no             & yes & no              & yes              \\
fraction of errors & 1     & $\sfrac{1}{8}$ & $\sfrac{7}{8}$ & $\sfrac{1}{8}$ & 1     & $\sfrac{1}{8}$ & $\sfrac{7}{8}$ & $\sfrac{1}{8}$ & 1   & $\sfrac{3}{24}$ & $\sfrac{21}{24}$ \\
contribute to      & $H$   & $H$            & $D$            & $L$            & $L$   & $L$            & $D$            & $L$            & $D$ & $L$             & $D$              \\ \hline
\multirowcell{4}{Update rules\\to add this block} 
& \multicolumn{11}{l}{$H\leftarrow H\left(P_0(p_{cx}, n+2) + \sfrac{1}{8}P_1(p_{cx}, n+2)\right)$}                               \\
& \multicolumn{11}{l}{$L\leftarrow LP_0(p_{cx}, n+2) + \sfrac{1}{8}(L+E)P_1(p_{cx}, n+2) + \sfrac{1}{8}(H+L+E)P_2(p_{cx}, n+2)$} \\
& \multicolumn{11}{l}{$E\leftarrow  0$}                                                                                          \\
& \multicolumn{11}{l}{$D\leftarrow D+ EP_0(p_{cx}, n+2) + \sfrac{7}{8}(H+L+E)\left(1 - P_0(p_{cx}, n+2)\right)$} \\ \hline
\end{tabular}
    \caption{Classification of errors in the final measurement block.}
    \label{tab:fin}
\end{table*}

\end{document}